\documentclass[aps,prb,reprint,superscriptaddress,nofootinbib,longbibliography,floatfix,amsmath,amssymb]{revtex4-2}

\usepackage{graphicx}
\graphicspath{{../}{../figure/}{figure/}}
\usepackage{dcolumn}
\usepackage{bm}
\usepackage{booktabs}
\usepackage{xcolor}
\usepackage{microtype}
\usepackage[colorlinks=true, linkcolor=blue, citecolor=blue, urlcolor=blue, breaklinks=true]{hyperref}
\hypersetup{
  pdftitle={Supersymmetric quantum criticality with discrete symmetry},
  pdfauthor={Teng-Yue Wang and Shuai Yin}
}

\begin{document}
\title{Supersymmetric quantum criticality with discrete symmetry}
\author{Teng-Yue Wang}
\affiliation{School of Physics, Sun Yat-sen University, Guangzhou 510275, China}
\affiliation{Guangdong Provincial Key Laboratory of Magnetoelectric Physics and Devices, Sun Yat-sen University, Guangzhou 510275, China}
\author{Shuai Yin}
\email{yinsh6@mail.sysu.edu.cn}
\affiliation{School of Physics, Sun Yat-sen University, Guangzhou 510275, China}
\affiliation{Guangdong Provincial Key Laboratory of Magnetoelectric Physics and Devices, Sun Yat-sen University, Guangzhou 510275, China}

\date{May 30, 2026}

\begin{abstract}
Supersymmetry, originally proposed in high-energy physics, can emerge as a remarkable low-energy structure in condensed matter systems. While emergent supersymmetry at quantum critical points is widely discussed in models with continuous symmetries, real materials are constrained by microscopic discrete symmetries. To address this, we investigate $(2+1)$-dimensional Gross-Neveu-Yukawa theories coupling Dirac fermions to a complex order parameter with discrete $Z_n$ anisotropy. Using the functional renormalization group, we find that for $n>3$, the anisotropic perturbations are irrelevant at the fixed point, yielding a $\mathcal{N}=2$ Wess-Zumino supersymmetric critical point. In the ordered phase, this dangerously irrelevant anisotropy gives rise to a second characteristic length scale, $\xi'$, alongside the usual correlation length, $\xi$. By tracking mass thresholds along symmetry-broken renormalization group trajectories, we extract the exponents $\nu'$ and $\nu$ without imposing prior scaling assumptions. For the $Z_4$, $Z_5$, and $Z_6$ models, our results support the scaling relation $\nu'/\nu = 1 + |y_n|/p$ with $p=2$ in the isotropic framework used here.
\end{abstract}

\maketitle

\section{Introduction}\label{sec:introduction}

Supersymmetry (SUSY) was initially proposed in particle physics to address profound problems such as the hierarchy problem, where the symmetry between fermions and bosons leads to cancellations of quantum corrections \cite{weinberg2000quantum, wess1992supersymmetry, nillesSupersymmetrySupergravityParticle1984, haberSearchSupersymmetryProbing1985, wessSupergaugeTransformationsFour1974, GERVAIS1971632, dimopoulosSoftlyBrokenSupersymmetry1981}. However, direct experimental evidence for this elegant theoretical framework, such as supersymmetric partners, remains elusive.

Intriguingly, condensed matter systems provide a distinct route for exploring SUSY as an emergent low-energy structure \cite{groverQuantumCriticalityTopological2012, groverEmergentSpacetimeSupersymmetry2014, ponteEmergenceSupersymmetrySurface2014, jianEmergentSpacetimeSupersymmetry2015, liObservationEmergentSpacetime2018, zerfSuperconductingQuantumCriticality2016, yinFermioninducedQuantumCritical2020, huijseEmergentSupersymmetryIsing2015, jianEmergenceSupersymmetricQuantum2017, liEdgeQuantumCriticality2017, yuFinitescaleEmergence22019, yuEmergentSpacetimeSupersymmetry2022, liUncoveringEmergentSpacetime2024, fendleyLatticeModelsN22003, yuSimulatingWessZuminoSupersymmetry2010, bauerSupersymmetricMulticriticalPoint2013, rahmaniEmergentSupersymmetryStrongly2015, liSupersymmetryInteractingMajorana2019, maRealizationSupersymmetryIts2021, kavirajParisiSourlasSupersymmetryRandom2022,zengNonequilibriumCriticalDynamics2025}. A particularly relevant class of examples arises at quantum critical points, where bosonic and fermionic degrees of freedom can become related by an emergent spacetime SUSY \cite{groverQuantumCriticalityTopological2012, groverEmergentSpacetimeSupersymmetry2014, ponteEmergenceSupersymmetrySurface2014, jianEmergentSpacetimeSupersymmetry2015, liObservationEmergentSpacetime2018, zerfSuperconductingQuantumCriticality2016, yinFermioninducedQuantumCritical2020, huijseEmergentSupersymmetryIsing2015, jianEmergenceSupersymmetricQuantum2017, liEdgeQuantumCriticality2017, yuFinitescaleEmergence22019, yuEmergentSpacetimeSupersymmetry2022, liUncoveringEmergentSpacetime2024}. Prominent among such examples are $(2+1)$-dimensional systems of Dirac fermions coupled to a complex bosonic order parameter \cite{groverEmergentSpacetimeSupersymmetry2014, ponteEmergenceSupersymmetrySurface2014, jianEmergentSpacetimeSupersymmetry2015, zerfSuperconductingQuantumCriticality2016, liObservationEmergentSpacetime2018}, whose critical behavior is governed by the $(2+1)$-dimensional $\mathcal{N}=2$ Wess-Zumino fixed point, described by a superconformal field theory (SCFT) \cite{AHARONY199767, leeEmergenceSupersymmetryCritical2007}. Existing studies of this fixed point predominantly assume continuous symmetries, whereas real crystalline materials are fundamentally constrained by discrete lattice symmetries.

Discrete symmetries also enrich universal critical phenomena. Specifically, in $Z_n$ spin systems with sufficiently large $n$, the corresponding anisotropies can act as dangerously irrelevant couplings, which flow to zero at the critical point yet remain essential for the low-energy physics of the ordered phase \cite{fisher1975renormalization, nelsonCoexistencecurveSingularitiesIsotropic1976, AMIT1982207, leonardCriticalExponentsCan2015, okuboScalingRelationDangerously2015}. This allows an emergent continuous symmetry at the critical point. However, in the symmetry-broken ordered phase, the dangerously irrelevant anisotropy eventually freezes the angular fluctuations of the order parameter, leading to discrete symmetry breaking.

A hallmark of this dangerously irrelevant anisotropy is the emergence of a second characteristic length scale in the ordered phase. Unlike genuine continuous symmetry breaking, such discrete symmetry breaking does not produce true Goldstone modes. Instead, both longitudinal and transverse fluctuations acquire finite masses, giving rise to the two scales $\xi \sim |r|^{-\nu}$ and $\xi' \sim |r|^{-\nu'}$ \cite{AMIT1982207, leonardCriticalExponentsCan2015}, where $r$ denotes the distance to the critical point in parameter space. The second scale is controlled by the dangerous irrelevant coupling and is associated with an additional critical exponent $\nu'$. A scaling relation of the form $\nu'/\nu = 1+|y|/p$, where $y$ is the anisotropy exponent characterizing the discrete perturbation, has been widely discussed \cite{chubukovTheoryTwodimensionalQuantum1994, oshikawaOrderedPhaseScaling2000, louEmergenceU1Symmetry2007, leonardCriticalExponentsCan2015, okuboScalingRelationDangerously2015, shaoQuantumCriticalityTwo2016, patilUnconventionalU1Zq2021}. Studies on continuous-to-discrete crossovers in classical and quantum clock models suggest that $p$ typically takes the value $p=2$ in isotropic three-dimensional classical systems, while $p=3$ has been observed in $(2+1)$-dimensional quantum systems and attributed to spacetime anisotropy \cite{patilUnconventionalU1Zq2021}.

The impact of such discrete symmetries extends naturally to systems with fermions. An important example of this is the $(2+1)$-dimensional quantum phase transition from a Dirac semimetal (DSM) to a valence bond solid (VBS) phase with $Z_3$ symmetry \cite{castronetoElectronicPropertiesGraphene2009, houElectronFractionalizationTwodimensional2007, ryuMassesGraphenelikeTwodimensional2009, gutierrezImagingChiralSymmetry2016}. In the pure bosonic case, the $Z_3$ cubic anisotropy drives the phase transition to be first-order. However, for a sufficiently large number of fermion flavors $N_f$, gapless fermion fluctuations can soften this tendency and render the transition continuous. This mechanism gives rise to the fermion-induced quantum critical point (FIQCP) \cite{classenFluctuationinducedContinuousTransition2017, jianFermioninducedQuantumCritical2017a, liFermioninducedQuantumCritical2017, torresFermioninducedQuantumCriticality2018, liFermioninducedQuantumCritical2020, jianFermioninducedQuantumCritical2017}. Despite these novel properties, emergent SUSY is absent at this critical point.

Building on this background, a sequence of central questions arises: can $(2+1)$-dimensional $\mathcal{N}=2$ SUSY emerge in systems with discrete symmetries, such as $Z_n$ ($n>3$)? If so, how do these discrete symmetries enrich the critical behavior in the ordered phase? Specifically, will they act as dangerously irrelevant couplings to induce a second characteristic length scale, and, in the presence of gapless fermions, does the associated scaling law favor $p=2$ or $p=3$?

In this work, we investigate a class of $(2+1)$-dimensional Gross-Neveu-Yukawa (GNY) theories in which Dirac fermions are coupled to a complex bosonic order parameter with $Z_n$ ($n>3$) anisotropy. We find a fixed point where the anisotropic terms vanish as irrelevant perturbations, yielding an emergent $(2+1)$-dimensional $\mathcal{N}=2$ SUSY. This critical theory is described by the $\mathcal{N}=2$ Wess-Zumino model \cite{wessSupergaugeTransformationsFour1974, AHARONY199767, zerfSuperconductingQuantumCriticality2016}. Furthermore, we demonstrate this emergent SUSY through the relations among the couplings and the anomalous dimensions.

We then investigate how the discrete anisotropy enriches the critical behavior on the ordered side of this SUSY criticality. We follow the ordered-phase renormalization group (RG) trajectories and extract the characteristic scales $\xi$ and $\xi'$ from the thresholds of the longitudinal and transverse dimensionless masses ($m_{L/T}^2=1$). This allows us to determine the critical exponents $\nu$, $\nu'$, and their ratio $\nu'/\nu$ without assuming the scaling law in advance. Even in the presence of gapless fermions, for the $Z_4$, $Z_5$, and $Z_6$ models, the resulting exponent ratios satisfy $\nu'/\nu = 1 + |y_n|/2$, favoring the scaling law with $p=2$ rather than $p=3$.

We employ the functional renormalization group (FRG) method \cite{DUPUIS20211, BERGES2002223}, which captures non-perturbative effects through exact flow equations for the effective action. This approach is well suited for determining critical exponents and tracking RG trajectories in strongly coupled systems. Moreover, it has been shown to be a powerful tool for studying coupled fermion-boson systems \cite{janssenAntiferromagneticCriticalPoint2014, giesFunctionalPerspectiveEmergent2017, feldmannCriticalWessZuminoModels2018}.

This paper is structured as follows. In Sec.~\ref{sec:model}, we introduce the $Z_n$-symmetric GNY model. Section~\ref{sec:frg} details the FRG framework and flow equations. Numerical results on the SUSY critical point, ordered-phase RG flow, two length scales, and the scaling law are presented in Sec.~\ref{sec:results}. We discuss the interpretation of the scaling law and compare different estimates of $\nu'$ in Sec.~\ref{sec:discussion}.

\section{Model}\label{sec:model}

Dirac semimetals \cite{vafekDiracFermionsSolids2014, wehlingDiracMaterials2014} are a class of materials characterized by linear dispersion relations, where low-energy excitations can be described by the Dirac equation. In two dimensions, the prototypical example is graphene \cite{novoselovTwodimensionalGasMassless2005}, which consists of a hexagonal lattice of carbon atoms. The low-energy physics of graphene is primarily influenced by two inequivalent Dirac points, known as the $K$ and $K'$ points \cite{castronetoElectronicPropertiesGraphene2009}. In this material, the low-energy excitations emerge as massless four-component Dirac fermions, which correspond to both sublattice and valley degrees of freedom \cite{semenoffCondensedMatterSimulationThreeDimensional1984}.

In graphene, electronic interactions can drive ordered phases that open a gap in the Dirac spectrum \cite{herbutInteractionsPhaseTransitionsGraphene2006, honerkampDensityWavesCooper2008, herbutTheoryInteractingElectrons2009, weeksInteractiondrivenInstabilitiesDirac2010}. The Kekul\'e VBS order provides a representative example: it corresponds to a modulation of the nearest-neighbor hopping amplitudes, reduces the $C_6$ symmetry to a $C_3$ one, and thereby realizes a $Z_3$-ordered phase \cite{ryuMassesGraphenelikeTwodimensional2009, royUnconventionalSuperconductivityHoneycomb2010, houElectronFractionalizationTwodimensional2007, gomesDesignerDiracFermions2012, gutierrezImagingChiralSymmetry2016}. This mechanism motivates a broader class of Dirac systems in which a complex order parameter is subject to a discrete $Z_n$ anisotropy. At the field-theory level, this corresponds to coupling a complex order parameter with $Z_n$ anisotropy, analogous to that of an $n$-state clock model \cite{hasenbuschAnisotropicPerturbationsThreedimensional2011, hasenbuschMonteCarloStudy2019, louEmergenceU1Symmetry2007, oshikawaOrderedPhaseScaling2000, todorokiOrderedPhasePhase2002, zhitomirskyNatureFinitetemperatureTransition2014, leonardCriticalExponentsCan2015, okuboScalingRelationDangerously2015, patilUnconventionalU1Zq2021}, to Dirac fermions.

The corresponding $n$-state order parameter is represented by the complex scalar field $\phi = (\phi_1 + \mathrm{i} \phi_2)/\sqrt{2}$ \cite{ryuMassesGraphenelikeTwodimensional2009}. For $Z_3$, the symmetry allows the existence of cubic terms, leading to a first-order phase transition within the Landau-Ginzburg-Wilson paradigm \cite{golnerInvestigationPottsModel1973, wuPottsModel1982}. Fermionic fluctuations can significantly alter this picture, causing it to become continuous for a sufficiently large number of fermion flavors $N_f$, i.e., a FIQCP \cite{schererGaugefieldassistedKekuleQuantum2016, liFermioninducedQuantumCritical2017, jianFermioninducedQuantumCritical2017a}. However, for the $N_f=0.5$ case where the bosonic and fermionic degrees of freedom are matched, the $Z_3$ perturbation is relevant and emergent SUSY is absent. Therefore, we focus on $Z_n$ models with $n>3$.

To describe the low-energy effective theory of this system, we employ the GNY model. We construct the corresponding Lagrangian from three components: the boson component $\mathcal{L}_{\phi}$, which encodes the dynamics and self-interactions of the order parameter field; the fermion component $\mathcal{L}_{\psi}$, which describes massless Dirac fermions; and the Yukawa coupling term $\mathcal{L}_{\psi\phi}$, which accounts for the interactions between fermions and bosons. In the absence of anisotropy, we can effectively map this model to the supersymmetric $\mathcal{N}=2$ Wess-Zumino model \cite{wessSupergaugeTransformationsFour1974, AHARONY199767, zerfSuperconductingQuantumCriticality2016} by carefully fine-tuning the coupling constants and the number of fermion flavors. In the following, we discuss each of these components.

\subsection{Boson Component}
For the bosonic field $\phi = (\phi_1 + \mathrm{i} \phi_2)/\sqrt{2}$ introduced above, an isotropic potential would possess a continuous $U(1)$ rotational symmetry. Here, we incorporate discrete microscopic symmetry by introducing an anisotropic potential $V(\phi, \phi^*)$ permitted by $Z_n$ symmetry. The general form of the bosonic Lagrangian is thus given by
\begin{equation}
    \mathcal{L}_\phi = -\phi^* \partial^2 \phi + V(\phi, \phi^*).
\end{equation}

To construct $V(\phi, \phi^*)$, we define the $Z_n$ invariants as:
\begin{equation}
\begin{aligned}
\overline{\rho} &= |\phi|^2=\frac{1}{2}(\phi_1^2+\phi_2^2),\\
\overline{\tau}^\prime_3 &= \frac{1}{\sqrt{2}}(\phi_1^3-3\phi_1\phi_2^2),\\
\overline{\tau}^\prime_4 &= \frac{1}{2}(\phi_1^4-6\phi_1^2\phi_2^2+ \phi_2^4),\\
\overline{\tau}^\prime_5 &= \frac{\sqrt{2}}{4}(\phi_1^5-10\phi_1^3\phi_2^2+ 5\phi_1\phi_2^4),\\
\overline{\tau}^\prime_6 &= \frac{1}{4}(\phi_1^6-15\phi_1^4\phi_2^2+15\phi_1^2\phi_2^4-\phi_2^6).
\end{aligned}
\end{equation}
Here, $\overline{\tau}_n^\prime = \phi^n + \phi^{*n}=2\,\operatorname{Re}(\phi^n)$.

For a complex scalar variable $\phi$ with two real degrees of freedom, the invariant ring under $Z_n$ is generated by three invariants:
\begin{equation}
    \begin{aligned}
        |\phi|^2,\quad (\phi^n+\phi^{*n}),\quad \mathrm{i}(\phi^n-\phi^{*n}),
    \end{aligned}
\end{equation}
subject to the relation $(\phi^n+\phi^{*n})^2 - (\phi^n-\phi^{*n})^2 = 4|\phi|^{2n}$. 
In this work, we further restrict the anisotropic potential to be invariant under the reflection $\phi\to\phi^*$, which renders $\mathrm{i}(\phi^n-\phi^{*n})$ odd and therefore excludes it from the Landau potential. This choice is physically motivated, for example, by the $Z_3$ Kekul\'e order, where the combined action of sublattice symmetry and time-reversal symmetry acts as such a reflection on the complex order parameter \cite{ryuMassesGraphenelikeTwodimensional2009, torresFermioninducedQuantumCriticality2018}. As a result, only $\overline{\rho}$ and $\overline{\tau}_n^\prime$ are retained, meaning the general symmetry-allowed potential is constructed as a function of these two invariants, $V(\phi, \phi^*) = V(\overline{\rho}, \overline{\tau}_n^\prime)$.

\subsection{Fermion Component}
In a $(2+1)$-dimensional Euclidean metric, the Lagrangian for the fermionic kinetic term is \cite{semenoffCondensedMatterSimulationThreeDimensional1984}
\begin{equation}
    \begin{aligned}
        \mathcal{L}_\psi &= \overline{\psi} \gamma_\mu \partial_\mu \psi, \quad \mu = 0,1,2 ,\\
\gamma_0 &= \mathbb{I}_2\otimes \sigma_z ,\quad \gamma_1 = \sigma_z \otimes \sigma_y,\quad \gamma_2 = \mathbb{I}_2\otimes \sigma_x ,
    \end{aligned}
\end{equation} 
where $\psi = \bigoplus_{i=1}^{N_f}\psi_i$, and each $\psi_i$ is a four-component spinor. The $\gamma$ matrices are composed of Pauli matrices and satisfy the Clifford algebra $\{\gamma_\mu, \gamma_\nu\} = 2\delta_{\mu\nu}$. We define $d_\gamma=4$ as the dimension of the gamma matrices.

The Dirac fermion Lagrangian describes the dynamics of massless fermions. $N_f$ fermion flavors, or equivalently, $N_f$ pairs of Dirac points are introduced. For example, $N_f=2$ corresponds to spin-1/2 electrons (e.g., in graphene).

\subsection{Yukawa Coupling Term}
The coupling between fermions and the order parameter field $\phi = (\phi_1 + \mathrm{i} \phi_2)/\sqrt{2}$ is \cite{royUnconventionalSuperconductivityHoneycomb2010}:

\begin{equation}
\mathcal{L}_{\psi\phi} = \mathrm{i}\overline{h} \overline{\psi} \left( \phi_1 \gamma_3 + \phi_2 \gamma_5 \right) \psi .
\end{equation}
Here, we define the remaining gamma matrices:
\begin{equation}
    \begin{aligned}
        \gamma_3 &= \sigma_x \otimes \sigma_y, \\
\gamma_5 &= \sigma_y \otimes \sigma_y, \\
\gamma_{35} &= -\mathrm{i} \gamma_3\gamma_5.
    \end{aligned}
\end{equation}
Under the transformation $\psi \to e^{\mathrm{i}\frac{\theta\gamma_{35}}{2}}\psi,\phi \to e^{\mathrm{i}\theta}\phi$, the Yukawa term possesses a $U(1)$ symmetry.

\subsection{Total Lagrangian}

Finally, the total Lagrangian is
\begin{equation}
\begin{aligned}
\mathcal{L}=\mathcal{L}_\psi+\mathcal{L}_{\psi\phi}+\mathcal{L}_\phi.
\end{aligned}
\end{equation}
Here, both the fermion and Yukawa components exhibit $U(1)$ rotational symmetry, while the boson component with $Z_n$ anisotropy may break it.

Although generic quantum materials exhibit spacetime anisotropy, we employ an isotropic Euclidean spacetime ansatz (setting the fermionic and bosonic velocities to one) in our effective action. This isotropic setup is physically motivated by the established RG result that Lorentz symmetry emerges at the critical fixed point for this class of Yukawa theories \cite{royEmergentLorentzSymmetry2015}. We will proceed with this formalism to extract the universal critical behavior, leaving the discussion on its limitations and the potential impact of spacetime anisotropy on specific scaling laws to Sec.~\ref{sec:discussion}.

\section{FRG Derivation}\label{sec:frg}

\subsection{FRG Method}

The functional renormalization group \cite{DUPUIS20211, BERGES2002223} builds on the idea of Wilson's renormalization group, which consists of successively integrating out high-energy degrees of freedom to access the infrared properties of a system. The key difference is that FRG deals with functionals of fields rather than a finite set of coupling constants.

Consider the partition function:
\begin{equation}
    Z[J] = \int \mathcal{D}\varphi\,
    \exp\left[-S[\varphi]+\int_{\mathbf r} J\cdot\varphi\right].
\end{equation}

To gradually incorporate infrared fluctuations as the scale $k$ is lowered, FRG introduces a scale-dependent mass-like regulator $R_k(p)$. The corresponding regulator term reads
\begin{equation}
    \Delta S_k = \frac{1}{2}\int_p\varphi(-p) R_k (p)\varphi(p).
\end{equation}

Adding this regulator term to the original action yields a scale-dependent partition function $Z_k[J]$. The corresponding effective action is then defined via a modified Legendre transformation
\begin{equation}
    \Gamma_k[\Phi]
    = -W_k[J] + \int_{\mathbf r} J\cdot\Phi - \Delta S_k[\Phi],
\end{equation}
where $W_k[J]=\ln Z_k[J]$ and $\Phi=\delta W_k[J]/\delta J$.

In the limit $k\to \Lambda$ (the ultraviolet cutoff), $R_k \to \infty$, thereby suppressing all fluctuations. As $k\to 0$, $R_k \to 0$, so that all fluctuations are restored and $\Gamma_{k=0}$ contains the complete information of the model. The evolution of $\Gamma_k$ with $k$ is governed by an exact functional differential equation, the Wetterich equation \cite{wetterichExactEvolutionEquation1993, DUPUIS20211, BERGES2002223}:
\begin{equation}\label{eq-wetterich}
    \partial_t \Gamma_k = \frac{1}{2} \mathrm{STr} \left[ 
    \left( \Gamma_k^{(2)} + R_k \right)^{-1} \partial_t R_k 
\right],
\end{equation}
where $t$ is defined as $\ln\left(\frac{k}{\Lambda}\right)$. Solving this equation yields the exact renormalization group flow.

Exact solutions of the Wetterich equation are generally intractable, making approximations necessary. However, even at the lowest order of approximation, FRG can capture non-perturbative properties of the system. Here we employ the improved local potential approximation (LPA$'$) \cite{DUPUIS20211}, which retains anomalous dimensions but neglects the momentum dependence of higher-order vertices, treating them as a local effective potential.

\subsection{Effective Action}

Within the LPA$'$ truncation, the scale-dependent effective action for the present model is taken as
\begin{equation}
\begin{aligned}
    \Gamma_k =& \int d^D x \left[ {Z_{\psi,k}} \overline{\psi}\gamma_\mu\partial_\mu\psi + \mathrm{i}{\overline{h}_k} \overline{\psi}(\phi_1\gamma_3+\phi_2\gamma_5)\psi \right.\\
    &\left.- \frac{1}{2}{Z_{\phi,k}}(\phi_1\partial^2\phi_1+\phi_2\partial^2\phi_2) + U_k({\overline{\rho}},{\overline{\tau}^\prime}) \right].
\end{aligned}
\end{equation}
In this ansatz, $Z_{\psi,k}$ and $Z_{\phi,k}$ are the scale-dependent wave-function renormalization factors for the fermionic and bosonic fields, respectively. Their logarithmic flows define the anomalous dimensions,
\begin{align}
\eta_\phi &= -\partial_t \ln{Z_{\phi,k}}, \\
\eta_\psi &= -\partial_t \ln{Z_{\psi,k}}.
\end{align}

Following the above notation, the dimensionless effective potential $u$ is defined as $u(\rho,\tau_n^\prime) = k^{-D}U_k(\overline{\rho},\overline{\tau}_n^\prime)$, with
\begin{equation}
\begin{aligned}
    \rho &= k^{2-D} Z_{\phi,k}\,\overline{\rho},\\
    \tau_n^\prime &= k^{-n(D-2)/2} Z_{\phi,k}^{n/2}\,
    \overline{\tau}_n^\prime, \qquad n=3,4,5,6 .
\end{aligned}
\end{equation}
Here, $\rho$ and $\tau_n^\prime$ are dimensionless invariants, while $\overline{\rho}$ and $\overline{\tau}_n^\prime$ denote the corresponding dimensionful invariants. The Yukawa coupling is rescaled in the same convention as
\begin{equation}
    h^2 = k^{D-4} Z_{\phi,k}^{-1} Z_{\psi,k}^{-2} \overline{h}_k^2 .
\end{equation}

In the ordered phase, the effective potential is expanded around one of its running minima, $\kappa=\rho_0$. For a $Z_n$-symmetric potential, the $n$ minima are equivalent, and one of them can be chosen as $\phi_1^{\min} = -\sqrt{2\kappa}, \, \phi_2^{\min} = 0$. Extending the treatment of the $Z_3$ system in Ref.~\cite{torresFermioninducedQuantumCriticality2018}, we redefine the anisotropic invariants $\tau_n^\prime$ into $\tau_n$ such that $\tau_n$ vanishes at this minimum:
\begin{equation}
    \begin{aligned}
        \tau_3 &=  2\rho^\frac{3}{2} +\tau_3^\prime,\\
        \tau_4 &= 2\rho^{2} - \tau_4^\prime ,\\
        \tau_5 &=  2\rho^\frac{5}{2} +\tau_5^\prime,\\
        \tau_6 &= 2\rho^{3} - \tau_6^\prime .\\
    \end{aligned}
\end{equation}
This redefinition simplifies the projection of the flow equations around $\kappa$.

To perform practical calculations within the FRG framework, the effective potential must be truncated to a finite order.
 Research on $Z_3$ FIQCP indicates that a truncation to $\phi^4$ order is insufficient to capture all the relevant physics in such models, whereas results obtained from truncations at $\phi^6$ order and beyond are mutually consistent, as higher-order terms are irrelevant \cite{classenFluctuationinducedContinuousTransition2017}. The same argument applies to general $Z_n$ symmetries. Therefore, in this study, we consistently truncate the boson component to $\phi^6$ order.

\subsection{Flow Equations}

To evaluate the Wetterich equation Eq.~\eqref{eq-wetterich}, the Hessian $\Gamma_k^{(2)}$ and the regulator matrix $R_k$ entering the supertrace must be specified. Following the convention of Ref.~\cite{BERGES2002223}, these matrices are written in the field basis $(\phi_1,\phi_2,\overline{\psi},\psi)$. The Hessian takes the form
\begin{equation}\label{eqGamma(2)}
    \Gamma_k^{(2)} = \begin{pmatrix}
    \Gamma_{k,11}^{(2)} & \Gamma_{k,12}^{(2)} &\Gamma_{k,1\overline{F}}^{(2)} & \Gamma_{k,1F}^{(2)} \\
\Gamma_{k,21}^{(2)} & \Gamma_{k,22}^{(2)} &\Gamma_{k,2\overline{F}}^{(2)} & \Gamma_{k,2F}^{(2)} \\
\Gamma_{k,\overline{F}1}^{(2)} & \Gamma_{k,\overline{F}2}^{(2)} &\Gamma_{k,\overline{F}\overline{F}}^{(2)} & \Gamma_{k,\overline{F}F}^{(2)} \\
\Gamma_{k,F1}^{(2)} & \Gamma_{k,F2}^{(2)} &\Gamma_{k,F\overline{F}}^{(2)} & \Gamma_{k,FF}^{(2)}
\end{pmatrix}.
\end{equation}
Here the subscripts $1,2,\overline{F},F$ denote derivatives with respect to
$\phi_1,\phi_2,\overline{\psi},\psi$, respectively. Left derivatives are used throughout; for example,
\begin{equation}
\Gamma_{k,11}^{(2)}=\frac{\delta^2 \Gamma_k}{\delta\phi_1\delta\phi_1},\quad
\Gamma_{k,1\overline{F}}^{(2)}=\frac{\delta^2 \Gamma_k}{\delta\phi_1\delta\overline{\psi}},\quad
\Gamma_{k,\overline{F}F}^{(2)}=-\frac{\delta^2 \Gamma_k}{\delta\overline{\psi}\delta\psi}.
\end{equation}

With the same convention, the regulator matrix is chosen to have the corresponding block structure,
\begin{equation}
    R_k = \begin{pmatrix}
    R_{\phi,k} & 0& 0 & 0 \\
0& R_{\phi,k} & 0 & 0 \\
0& 0 & 0 & R_{\psi,k} \\
0& 0 & -R_{\psi,k} & 0
\end{pmatrix},
\end{equation}
with
\begin{equation}
    R_{\psi,k} = \mathrm{i}Z_{\psi,k}q_\mu \gamma_\mu  r_\psi\left(q\right),
    \qquad
    R_{\phi,k} = Z_{\phi,k}q^2  r_\phi\left(q\right).
\end{equation}
The Litim regulator is used \cite{LITIM2002128, litimOptimisedRenormalisationGroup2001, litimMindGap2001, litimOptimisationExactRenormalisation2000}:
\begin{equation}
    \begin{aligned}
        r_{\psi}(q) &= \left(\frac{k}{q}-1\right)  \theta(k^{2} - q^{2}),\\
        r_{\phi}(q) &= \left( \frac{k^{2}}{q^{2}} - 1 \right) \theta(k^{2} - q^{2}).
    \end{aligned}
\end{equation}

The notation used in the following calculation follows Ref.~\cite{classenFluctuationinducedContinuousTransition2017}: derivatives with respect to the invariants are denoted by $u^{(n,m)}=\partial^{n+m}u/\partial\rho^n\partial\tau^m$, while derivatives with respect to field components are denoted by $u_{ij}=\partial^2u/\partial\phi_i\partial\phi_j$. Diagonalizing the bosonic block of the Hessian in Eq.~\eqref{eqGamma(2)} gives the longitudinal and transverse masses, $m_{L/T}^2=(u_{11}+u_{22})/2\pm\sqrt{4u_{12}^2+(u_{11}-u_{22})^2}/2$. Local three-point vertices at the minimum of the effective potential are denoted by $\omega_{ijk}=u_{ijk}|_{\phi_1=-\sqrt{2\kappa},\phi_2=0}$; for example, $\omega_{221}=\left.u_{221}\right|_{\phi_1=-\sqrt{2\kappa},\phi_2=0}$.

The flow equations are obtained by projecting Eq.~\eqref{eq-wetterich} onto the corresponding running couplings. The projection scheme follows the treatment of Ref.~\cite{torresFermioninducedQuantumCriticality2018}, while the present work extends the anisotropic part to general $Z_n$ invariants.

\subsubsection{Flow Equation for the Effective Potential}

By definition of the effective potential, the projection of Eq.~\eqref{eq-wetterich} onto a constant bosonic background gives
\begin{equation}
    V\partial_tU_k = \frac{1}{2} \mathrm{STr} \left[ 
    \left( \Gamma_k^{(2)} + R_k \right)^{-1} \partial_t R_k 
\right],
\end{equation}
where $V$ is the total volume.

The quantum correction to the flow equation for the effective potential has the same structure for different $Z_n$ anisotropies, while the canonical scaling part depends on $n$. The flow equation for a $Z_n$ invariant reads
\begin{equation}\label{eq:znuflom}
\begin{aligned}
    \partial_t u =& -D u + (D-2+\eta_\phi) \left( \rho u^{(1,0)} + \frac{n}{2}\tau u^{(0,1)} \right) \\
    &+ 2v_D \left(l_0^{(B_1)}  + l_0^{(B_2)}\right)  - 2v_D N_f d_\gamma l_0^{(F)},
\end{aligned}
\end{equation}
where $v_D^{-1}=2^{D+1}\pi^{D/2}\Gamma(D/2)$. For $n=3$, this expression reduces to the corresponding result for the $Z_3$ Dirac system in Ref.~\cite{torresFermioninducedQuantumCriticality2018}.

\subsubsection{Flow Equation for the Yukawa Coupling}

The Yukawa coupling is projected from the fermion-boson three-point vertex:
\begin{equation}
    \partial_t h = -\frac{\mathrm{i}}{N_f d_\gamma } \operatorname{Tr} \left[ \gamma_5  \frac{\delta}{\delta \Delta\phi_2}\frac{\delta}{\delta \psi}\frac{\delta}{\delta \overline{\psi}} \partial_t \Gamma_k  \right]_{\substack{\Delta\phi_i=0 \\ p=0}}.
\end{equation}
Evaluating this projection yields
\begin{equation}\label{h2flow}
    \begin{aligned}
        \partial_t h^2 =& (D - 4 + \eta_\phi + 2\eta_\psi) h^2
+ 8 v_D h^4 \left( l_{11}^{(FB_1)} - l_{11}^{(FB_2)}\right)\\
&- 16 v_D \sqrt{2\kappa} h^4 \omega_{221} l_{111}^{(FB_1 B_2)}.
    \end{aligned}
\end{equation}
The form of this flow is independent of the explicit choice of $n$; the $Z_n$ dependence enters through $\omega_{ijk}$ and $m_{L/T}^2$. For $n=3$, it reproduces the Yukawa flow equation reported in Ref.~\cite{torresFermioninducedQuantumCriticality2018}.

\subsubsection{Anomalous Dimensions}

The anomalous dimensions are obtained from the momentum dependence of the two-point functions. The wave-function renormalization flows are projected as
\begin{equation}
\partial_t Z_{\phi} = \left. \frac{\partial}{\partial p^2} \int \frac{d^D p_1}{(2\pi)^D} \frac{\delta}{\delta \phi_1(-p)} \frac{\delta}{\delta \phi_1(p_1)} \partial_t \Gamma_k \right|_{\substack{
\overline{\psi} = \psi = 0 \\
\Delta\phi_1 =\Delta\phi_2 = 0 \\
p = 0
}},
\end{equation}
\begin{equation}
\begin{split}
    \partial_t Z_{\psi} ={}&
    -\frac{\mathrm{i}}{N_f d_\gamma } \operatorname{Tr}
    \Biggl[
    \gamma_\mu \frac{\partial}{\partial p_\mu}
    \int \frac{d^D p_1}{(2\pi)^D}\\
    &\times
    \frac{\delta}{\delta \psi(p_1)}
    \frac{\delta}{\delta \overline{\psi}(p)}
    \partial_t \Gamma_k
    \Biggr]_{\substack{\Delta\phi_i=0 \\ p=0}} .
\end{split}
\end{equation}
Evaluating these projections gives
\begin{equation}\label{etaflow}
    \begin{aligned}
        \eta_{\phi}=&\frac{8v_D}{D}\left[\omega_{221}^2m_{22}^{(B_1B_2)}+N_fd_\gamma h^2\left(m_4^{(F)}+\omega_\psi m_2^{(F)}\right)\right],\\
        \eta_\psi =& \frac{8v_D}{D}h^2\left(m_{12}^{(FB_1)}+m_{12}^{(FB_2)}\right),
    \end{aligned}
\end{equation}
where $\omega_\psi = 2h^2\kappa$. For $n=3$, these expressions agree with the anomalous dimensions obtained in Ref.~\cite{torresFermioninducedQuantumCriticality2018}.

\section{Results}\label{sec:results}

Based on the flow equations obtained in Sec.~\ref{sec:frg}, we now present the numerical results. The analysis begins with the identification of the SUSY critical point for $Z_n$ models with $n>3$ and the characterization of its emergent SUSY. We then turn to the ordered-phase RG flow, where the dangerously irrelevant anisotropy gives rise to a second characteristic length scale. By extracting the corresponding RG scales directly from the symmetry-broken (SSB) flow, we determine $\nu$ and $\nu'$ and test the scaling law for $\nu'/\nu$.

\subsection{SUSY Critical Point}

In the following, we present the numerical evidence for emergent SUSY at the critical point under $Z_n$ ($n>3$) anisotropic perturbations. We analyze the fixed point governing this criticality, evaluate the relevance of the $Z_n$ couplings to determine whether such an emergence is allowed, and examine the realization of supersymmetry through the relations among couplings and the anomalous dimensions.

\subsubsection{Symmetric Expansion}

The fixed-point analysis in this subsection is carried out in the symmetric (SYM) expansion. To approximate the effective potential, a Taylor expansion is performed around the symmetric point $(\rho,\tau)=(0,0)$ and truncated at a finite order $N$ \cite{DUPUIS20211, torresFermioninducedQuantumCriticality2018, leonardCriticalExponentsCan2015}. For a $Z_n$ invariant, the expansion is written as
\begin{equation}\label{eqlambda}
    u(\rho,\tau) = \sum_{i+j=1}^{2i+nj=N} \frac{1}{i!  j!} \lambda_{i,j;k} \rho^i {\tau}^j .
\end{equation}

The flow equations for the potential vertices $\lambda_{i,j;k}$ are obtained by inserting Eq.~\eqref{eqlambda} into the effective-potential flow equation and expanding around $(\rho,\tau)=(0,0)$. Their explicit form is listed in Eq.~\eqref{partiallambda}. Together with Eqs.~\eqref{h2flow} and \eqref{etaflow}, this gives a closed set of coupled ordinary differential equations (ODEs) for the SYM expansion.

We obtain the fixed points by setting the beta functions---the right-hand sides of these ODEs---to zero and iteratively solving the resulting coupled equations using the Newton-Raphson method. This numerical root-finding procedure directly evaluates the stability matrix of the RG flow. At the converged fixed point, critical exponents are extracted from the eigenvalues of this stability matrix; in particular, the correlation-length exponent $\nu$ is given by the inverse of the largest positive eigenvalue.

\subsubsection{SUSY Critical Point}

We find three fixed points with zero anisotropy: the Nambu-Goldstone (NG) fixed point lies in the ordered phase, the Dirac semimetal (DSM) fixed point in the disordered phase, and the SUSY fixed point between them. The corresponding dimensionless couplings and anomalous dimensions at these fixed points are summarized in Table~\ref{tab:physical-fixed-points}.

\begin{table}[!htbp]
    \caption{Fixed-point values of the dimensionless couplings and anomalous dimensions of the physical fixed points at $N_f=0.5$ within the SYM LPA$'$6 truncation. The couplings $\lambda_{ij}$ are defined in Eq.~\eqref{eqlambda}.}
    \label{tab:physical-fixed-points}
    \begin{ruledtabular}
    \begin{tabular}{lcccc}
             & $\lambda_{10}$ & $\lambda_{20}$ & $\lambda_{30}$ & $\lambda_{01}$ \\
        NG   & $-0.1739$ & $3.5140$ & $26.1590$   & $0.0000$ \\
        SUSY & $-0.0494$ & $5.4259$ & $33.5757$   & $0.0000$ \\
        DSM  & $ 0.2492$ & $1.1253$ & $-184.9078$ & $0.0000$ \\
        \colrule
             & $h^2$ & $\eta_\phi$ & $\eta_\psi$ &  \\
        NG   & $0.0000$ & $0.0000$ & $0.0000$ &  \\
        SUSY & $4.7648$ & $0.3500$ & $0.3250$ &  \\
        DSM  & $6.5903$ & $0.5009$ & $0.2495$ &  \\
    \end{tabular}
    \end{ruledtabular}
\end{table}

The two non-SUSY fixed points, the NG and DSM fixed points, govern the RG flow in the ordered and disordered phases, respectively. The NG fixed point is inherited from the $U(1)$-symmetric theory obtained when the anisotropic couplings vanish. For discrete $Z_n$ symmetry, however, dangerously irrelevant anisotropic couplings eventually freeze the would-be $U(1)$ angular mode, so no genuine Goldstone mode exists in the true infrared ordered phase. The NG fixed point corresponds to the critical point of the pure bosonic $n$-state clock model \cite{leonardCriticalExponentsCan2015}, but the Yukawa coupling $h^2$ is relevant and renders it unstable. The DSM fixed point controls the long-range behavior on the disordered side; it has only one marginal direction and no relevant directions \cite{torresFermioninducedQuantumCriticality2018}.

The intermediate fixed point realizes a SUSY critical point for $n>3$. Our FRG calculations show that all $Z_n$ anisotropic couplings flow to zero at this fixed point, and as shown in Table~\ref{tab:eigenvalues}, the stability matrix possesses exactly one relevant direction, indicating that this fixed point governs a continuous phase transition. This emergent SUSY can be further verified by the relations among couplings and the anomalous dimensions analyzed below.

\begin{table}[!htbp]
    \caption{The two largest eigenvalues of the stability matrix at $N_f = 0.5$ within the SYM LPA$'$6 truncation.}
    \label{tab:eigenvalues}
    \begin{ruledtabular}
    \begin{tabular}{ccccc}
        & $Z_3$ & $Z_4$ & $Z_5$ & $Z_6$ \\
        $\theta_1$ & 1.0762 & 1.0762 & 1.0762 & 1.0762 \\
        $\theta_2$ & 0.5061 & -0.6826 & -1.0154 & -1.0154 \\
    \end{tabular}
    \end{ruledtabular}
\end{table}

We first examine the emergence of SUSY at the level of the effective Lagrangian. Once the $Z_n$ anisotropies vanish under the RG flow, the effective Lagrangian takes the form of the $\mathcal{N}=2$ Wess-Zumino model \cite{AHARONY199767}. SUSY in this model imposes relations among coupling constants, such as $h^2=\lambda_{20}$ \cite{leeEmergenceSupersymmetryCritical2007, groverEmergentSpacetimeSupersymmetry2014}. As shown in Table~\ref{tab:physical-fixed-points}, the fixed-point values are $h^2=4.7648$ and $\lambda_{20}=5.4259$, with a relative deviation of about $12\%$, which is reasonably consistent with the SUSY prediction given the simplicity of the LPA$'$ ansatz.

A second criterion for emergent SUSY is the equality of the bosonic and fermionic anomalous dimensions. SCFT predicts $\eta_\phi=\eta_\psi=1/3$ for the corresponding $\mathcal{N}=2$ supersymmetric fixed point \cite{AHARONY199767}, in agreement with previous numerical studies \cite{zerfSuperconductingQuantumCriticality2016, bobevBootstrappingThreeDimensional2015}. As summarized in Table~\ref{tab:critical-exponents}, our SYM LPA$'$6 calculation gives $\eta_\phi=0.3500$ and $\eta_\psi=0.3250$, close to the exact prediction. The values of $\eta_\phi$, $\eta_\psi$, and $\nu$ are the same for different $Z_n$ because they are governed by the emergent $U(1)$-symmetric sector; the distinction between different discrete symmetries is encoded in the anisotropy eigenvalue $y_n$.

\begin{table}[!htbp]
    \caption{Critical exponents at the SUSY fixed point for $N_f=0.5$ within the SYM LPA$'$6 truncation. The exponent $y_n$ denotes the leading eigenvalue associated with the corresponding $Z_n$ anisotropy.}
    \label{tab:critical-exponents}
    \begin{ruledtabular}
    \begin{tabular}{ccccc}
        & $\eta_\phi$ & $\eta_\psi$ & $\nu$ & $y_n$ \\
        $Z_3$ & 0.3500 & 0.3250 & 0.9292 & 0.5061 \\
        $Z_4$ & 0.3500 & 0.3250 & 0.9292 & -0.6826 \\
        $Z_5$ & 0.3500 & 0.3250 & 0.9292 & -4.3428 \\
        $Z_6$ & 0.3500 & 0.3250 & 0.9292 & -7.0017 \\
    \end{tabular}
    \end{ruledtabular}
\end{table}

An explanation for why emergent SUSY survives for $n>3$ but fails for $Z_3$ lies in the property of the $\mathcal{N}=2$ Wess-Zumino superconformal field theory. The scalar cubic operator $\phi^3$ has a protected scaling dimension of $2$ \cite{AHARONY199767, jianFermioninducedQuantumCritical2017a}. As explicitly argued in Ref.~\cite{jianFermioninducedQuantumCritical2017a}, since this dimension is smaller than the spacetime dimension $D=3$, the $Z_3$ cubic anisotropy acts as a relevant perturbation that prevents emergent SUSY. For $n>3$, however, higher-order discrete perturbations lack such protection, leaving open the possibility that they are irrelevant. Our calculations address this theoretical expectation: as shown in Table~\ref{tab:critical-exponents}, we find $y_3>0$, but $y_n<0$ for $n>3$, indicating that these higher-order perturbations are irrelevant and allow SUSY to emerge in the latter cases.

\subsection{RG Flow and Two Length Scales in the Ordered Phase}

Here, we examine how the discrete symmetry influences the critical behavior on the ordered side of the transition. We track the renormalization group flow in the ordered phase to observe the signatures of two characteristic length scales. From these results, we extract the corresponding critical exponents $\nu$ and $\nu'$ and use them to evaluate the associated scaling law.

\subsubsection{Symmetry-Broken Expansion}

To investigate the ordered side of the phase transition, it is more convenient to use a SSB expansion, in which the running minimum is kept explicitly. The effective potential is expanded around $(\rho,\tau)=(\kappa,0)$, with $\kappa=\kappa_k$ depending on the RG scale. For a $Z_n$ invariant, this expansion takes the form
\begin{equation}\label{eqLambda}
    u(\rho,\tau) = \Lambda_{0,1;k}\tau + \sum_{i+j=2}^{2i+nj=N} \frac{1}{i!  j!} \Lambda_{i,j;k} (\rho-\kappa_k)^i {\tau}^j .
\end{equation}
The minimum condition imposes $\Lambda_{1,0;k}=0$.

By diagonalizing the bosonic block of the Hessian in Eq.~(\ref{eqGamma(2)}) at $\kappa$, the longitudinal and transverse masses can be explicitly evaluated. Specifically, we find that these mass terms take the form
\begin{equation}\label{eq:ordered-mass-scaling}
m_L^2=2\kappa\Lambda_{20}, \qquad m_T^2=n^2\kappa^{\frac{n}{2}-1}\Lambda_{01}.
\end{equation}
Note that the transverse mass $m_T^2$ calculated in Ref.~\cite{leonardCriticalExponentsCan2015} is $1/2$ of the result obtained here, which arises from a different convention in defining the invariant $\tau$.

Substituting Eq.~\eqref{eqLambda} into the effective-potential flow equation gives the beta functions for the couplings $\Lambda_{i,j;k}$, while the condition $\partial_t\Lambda_{1,0;k}=0$ determines the flow of the running minimum $\kappa_k$. The explicit expressions are listed in Eqs.~\eqref{partialLambda} and \eqref{partialkappa}. Together with the flows of $h^2$, $\eta_\phi$, and $\eta_\psi$ in Eqs.~\eqref{h2flow} and \eqref{etaflow}, these equations form a closed system of ODEs.

It should be emphasized that the SSB expansion is not a trivial continuation of the SYM expansion. Taking the limit $\kappa\to0$ in the SSB flow equations does not recover the SYM flow equations, because the propagator denominators in the SSB expansion lack the term $u^{(1,0)}$, which leads to singular behavior as $\kappa$ approaches zero. Therefore, the SSB expansion is employed on the ordered side of the transition.

Within the SSB expansion, the critical fixed point and its stability matrix can also be evaluated. Instead of focusing on its static critical properties, we obtain the ordered-phase RG trajectories by numerically integrating the closed ODE system derived above, which allows direct tracking of the evolution of the mass scales deep in the ordered phase.

\subsubsection{RG Flow}
We use the $Z_4$ model with $N_f=0.5$ to illustrate the ordered-phase RG flow and the underlying mechanisms. Other $Z_n$ models with $n>3$ exhibit similar behavior, with the corresponding $Z_5$ and $Z_6$ results shown in Appendix~\ref{app:ordered-extra}.

\begin{figure*}[tbp]
    \centering
    \begin{tabular}{@{}cc@{}}
        \includegraphics[width=0.47\textwidth]{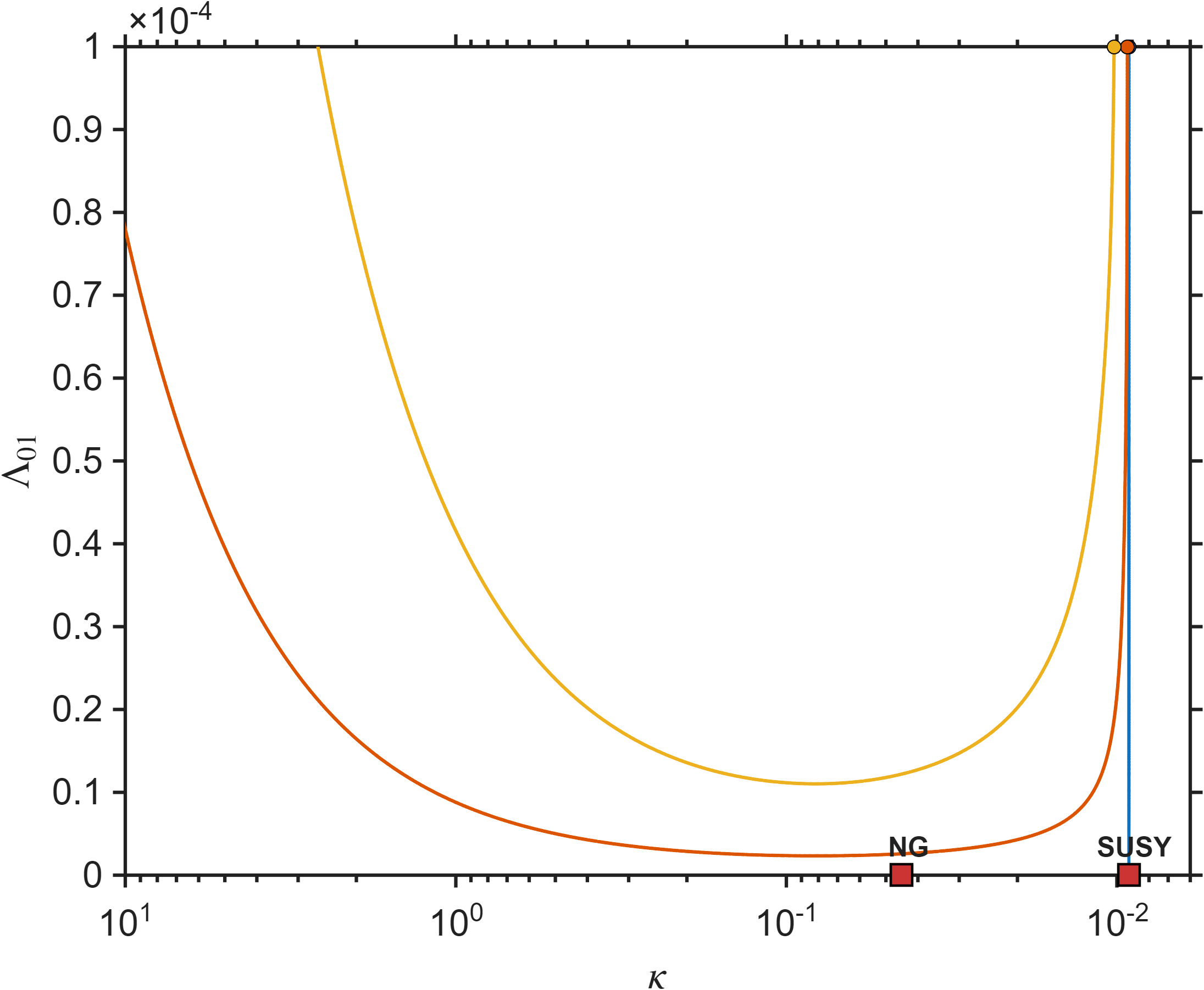} &
        \includegraphics[width=0.47\textwidth]{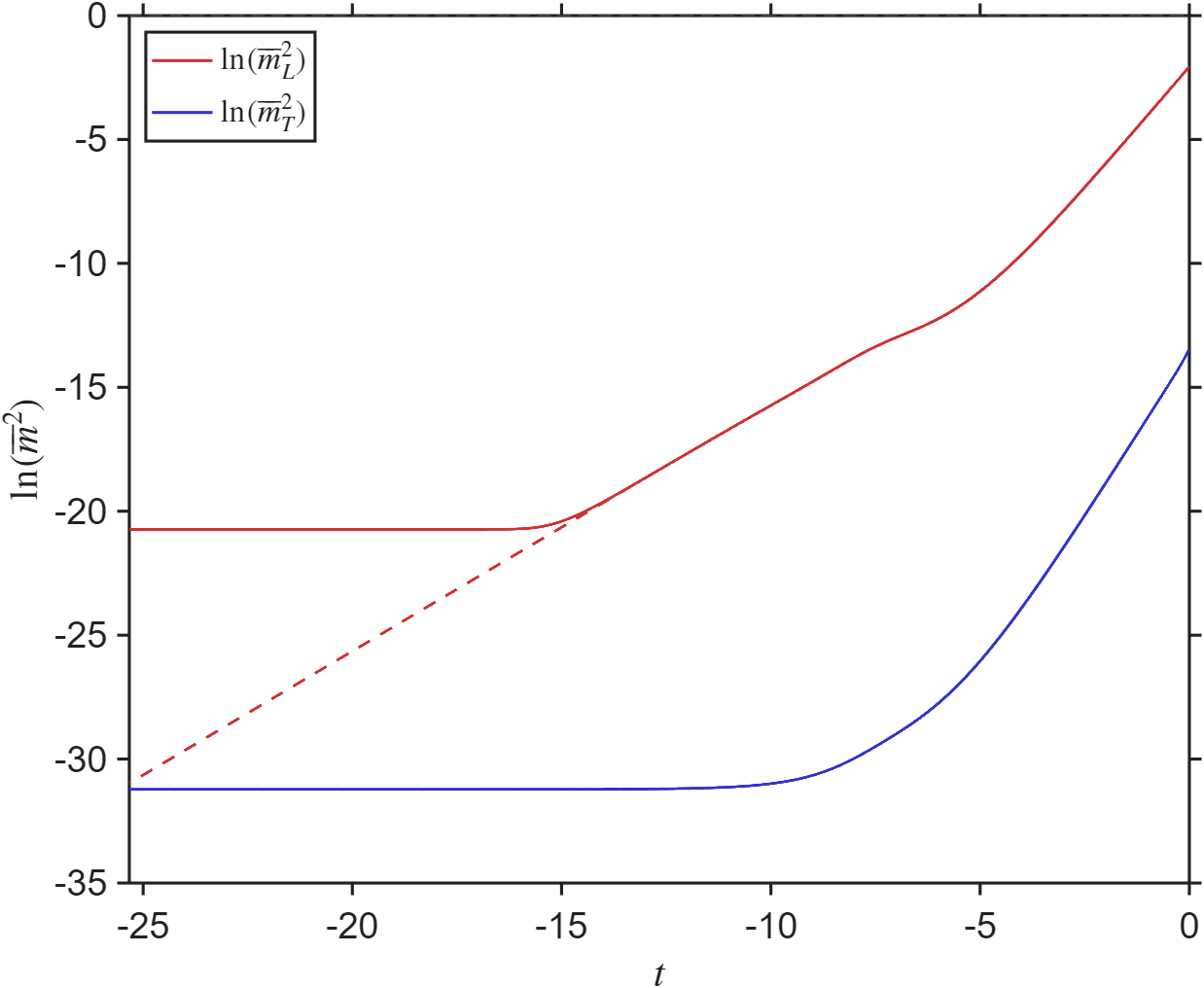} \\
        (a) & (b) \\[6pt]
        \includegraphics[width=0.47\textwidth]{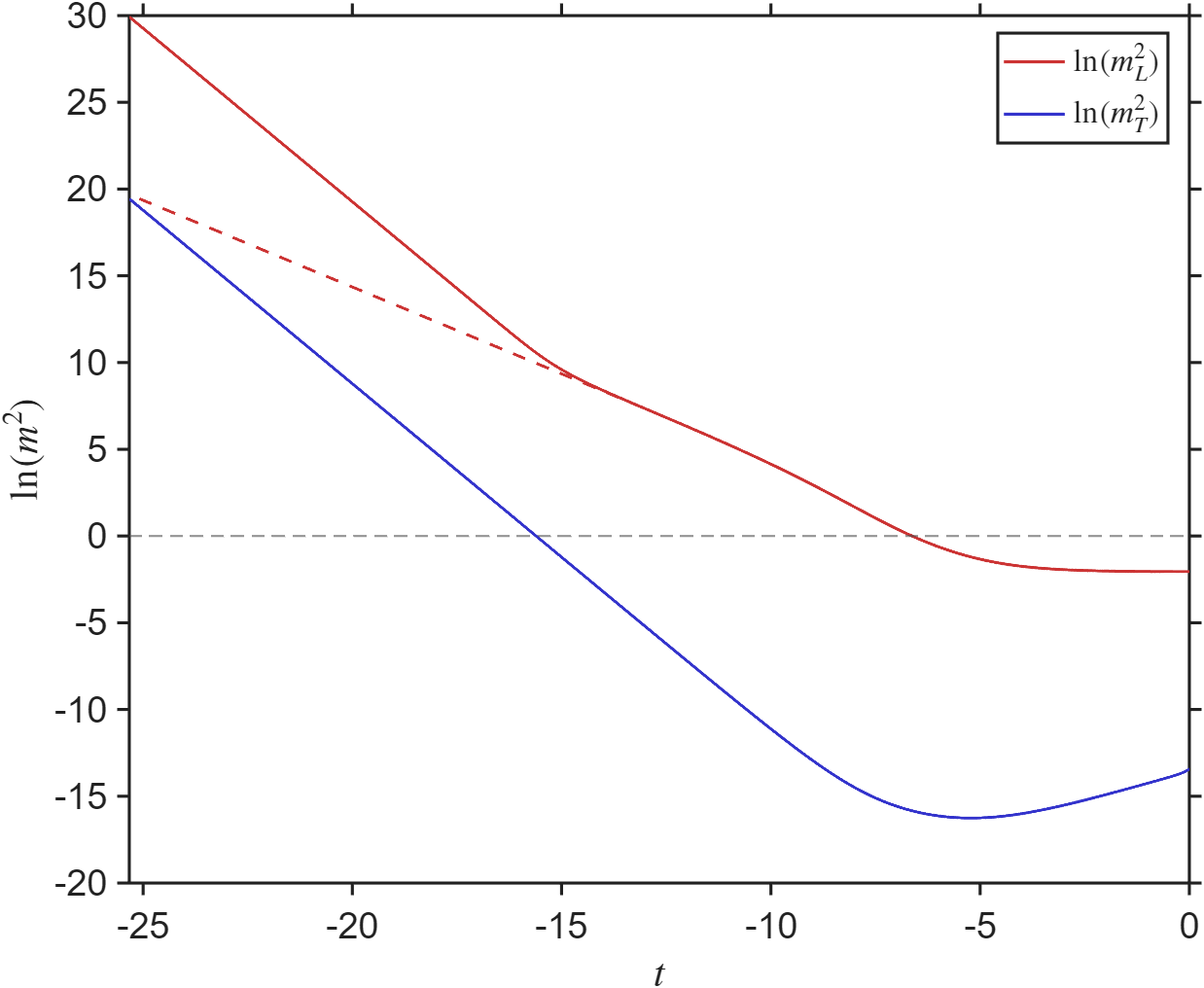} &
        \includegraphics[width=0.47\textwidth]{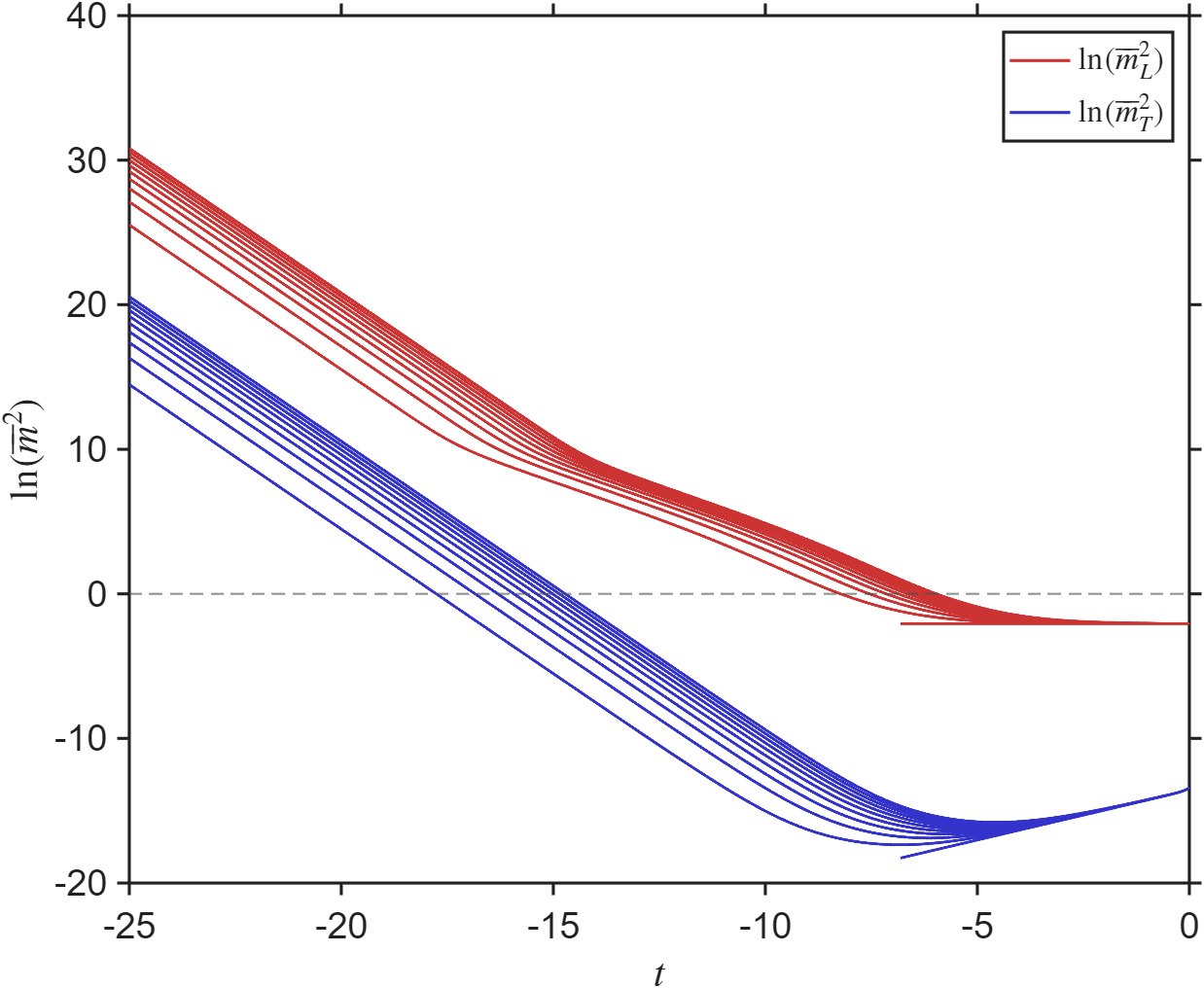} \\
        (c) & (d)
    \end{tabular}
    \caption{Ordered-phase RG flow and mass flows for the $Z_4$ model at $N_f=0.5$.
(a) Ordered-phase RG flow in the $(\kappa,\Lambda_{01})$ plane, illustrated by three representative trajectories.
(b) Flow of the dimensionful longitudinal mass $\ln \overline{m}_L^2$ (red) and transverse mass $\ln \overline{m}_T^2$ (blue) along a representative trajectory. For each mass, the dashed line corresponds to zero initial anisotropic coupling $\Lambda_{01}$, while the solid line corresponds to nonzero initial $\Lambda_{01}$.
(c) Flow of the corresponding dimensionless masses $\ln m_L^2$ (red) and $\ln m_T^2$ (blue), with the same line-style convention as in (b).
(d) Multi-trajectory dimensionless mass flows for different initial values of $\kappa$, used to determine the RG times $t_L$ and $t_T$ from the crossings $\ln m_L^2=0$ and $\ln m_T^2=0$.}
    \label{fig:rg-flows}
\end{figure*}

We first consider RG trajectories initialized near the SUSY fixed point with a small but nonzero anisotropic coupling $\Lambda_{01}$. By varying $\kappa$ on the ordered side, we observe the trajectories illustrated in Fig.~\ref{fig:rg-flows}(a). On this side, the trajectories first approach the NG fixed point and then bend away toward the anisotropic ordered regime. On the opposite side of the critical surface, the flow enters the disordered phase governed by the DSM fixed point, which is not shown here.

This bending of the trajectories reflects the structure of the ordered-phase RG flow. Near the critical region, the ordered-side flow initially approaches the NG fixed point, where anisotropic terms tend to vanish due to the dominance of the $U(1)$-symmetric sector. However, away from the critical point, the finite anisotropy gives the would-be Goldstone mode a small but nonzero transverse mass. As the RG scale is lowered, the corresponding dimensionless transverse mass grows, freezing the angular fluctuations and driving the system toward the anisotropic ordered regime deep in the ordered phase. This flow structure highlights the role of the NG fixed point in controlling an intermediate crossover regime within the ordered phase.

To explore the properties of the RG flow in greater detail, we restrict $\kappa$ to a narrow interval near $\kappa_c$ on the ordered side, with $\Lambda_{01}$ kept fixed and small. Based on the SSB flow equations, we compute and plot the flows of the dimensionful masses ($\overline{m}_{L/T}^2$) and the dimensionless masses ($m_{L/T}^2$), respectively, the former being related to inverse susceptibilities and the latter connected with the structure of the flow equations. The results are shown in Figs.~\ref{fig:rg-flows}(b) and \ref{fig:rg-flows}(c).

In Fig.~\ref{fig:rg-flows}(b), a key observation is that a nonzero anisotropic term significantly affects the renormalization of the longitudinal mass, preventing it from decreasing to zero and instead stabilizing it at a finite value. Similar to the $Z_3$ case \cite{torresFermioninducedQuantumCriticality2018}, this splitting of the mass flows due to discrete symmetry breaking introduces a second characteristic length scale into the system.

Fig.~\ref{fig:rg-flows}(c) shows the RG flows of dimensionless masses. A crucial observation is that the slopes of these curves exhibit clear changes at specific points: the slope of $\ln m_T^2$ changes significantly when $m_L^2 \approx 1$, while the slope of $\ln m_L^2$ changes when $m_T^2 \approx 1$.

As discussed in Ref.~\cite{leonardCriticalExponentsCan2015}, the points where $m_L^2=1$ and $m_T^2=1$ are associated with the inverse correlation lengths $\xi^{-1}$ and $\xi'^{-1}$, implying two characteristic length scales. This phenomenon finds a natural explanation in the structure of the flow equations. With $\Sigma_B$ in Eq.~\eqref{partialLambda}, the flow equations can be expressed in a form where the denominators contain terms like $(1+m_{L/T}^2)$. Consequently, the condition $m_{L/T}^2 = 1$ naturally defines a characteristic energy scale. These crossover points mark the thresholds where the respective mass terms become significant, causing a qualitative change in the RG flow.

\subsubsection{\texorpdfstring{Numerical Values of $\nu'$ and Scaling Law}{Numerical Values of nu prime and Scaling Law}}

As discussed in the previous section, the presence of dangerously irrelevant couplings in discrete symmetry models gives rise to two length scales. The second length scale, denoted as $\xi'$, introduces an additional exponent, $\nu'$, which plays a significant role in both theoretical and numerical studies \cite{leonardCriticalExponentsCan2015, 
AMIT1982207, nelsonCoexistencecurveSingularitiesIsotropic1976, okuboScalingRelationDangerously2015, patilUnconventionalU1Zq2021}.

However, since $\xi' > \xi$, the RG flow near the critical point is dominated by $\xi$. This dominance makes it challenging to extract the value of $\nu'$ associated with the second length scale from the stability matrix. Motivated by the relation between mass thresholds and length scales discussed in Ref.~\cite{leonardCriticalExponentsCan2015}, we therefore analyze the behavior of the entire RG flow. In the SSB flow, the point where $m_L^2 = 1$ corresponds to $\xi^{-1}$, while the point where $m_T^2 = 1$ corresponds to $\xi'^{-1}$. We denote the corresponding RG times by $t_L=t|_{\ln m_L^2=0}$ and $t_T=t|_{\ln m_T^2=0}$. Since $t_L=\ln(k_L/\Lambda)$ and $t_T=\ln(k_T/\Lambda)$, with $k_L\propto \xi^{-1}$ and $k_T\propto \xi^{\prime -1}$, these RG times differ from $\ln(\xi^{-1})$ and $\ln(\xi^{\prime -1})$ only by nonuniversal additive constants. In this way, the SSB RG flow provides a direct operational extraction of the two characteristic lengths in the present SUSY Dirac system.

At the SUSY critical point, there is only one relevant scaling direction. In the SSB calculation, we tune across the critical surface by varying the initial value of $\kappa$ while keeping the other couplings fixed at their fixed-point values. Although $\kappa$ need not coincide with the relevant scaling field, its deviation has a finite overlap with it; locally, this can be written as $u_r=c_\kappa(\kappa-\kappa_c)+O[(\kappa-\kappa_c)^2]$ with $c_\kappa\neq0$. Thus, $|\kappa-\kappa_c|$ tracks the amplitude of the relevant perturbation to leading order.

\begin{figure*}[tbp]
    \centering
    \begin{tabular}{ccc}
        \includegraphics[width=0.31\textwidth]{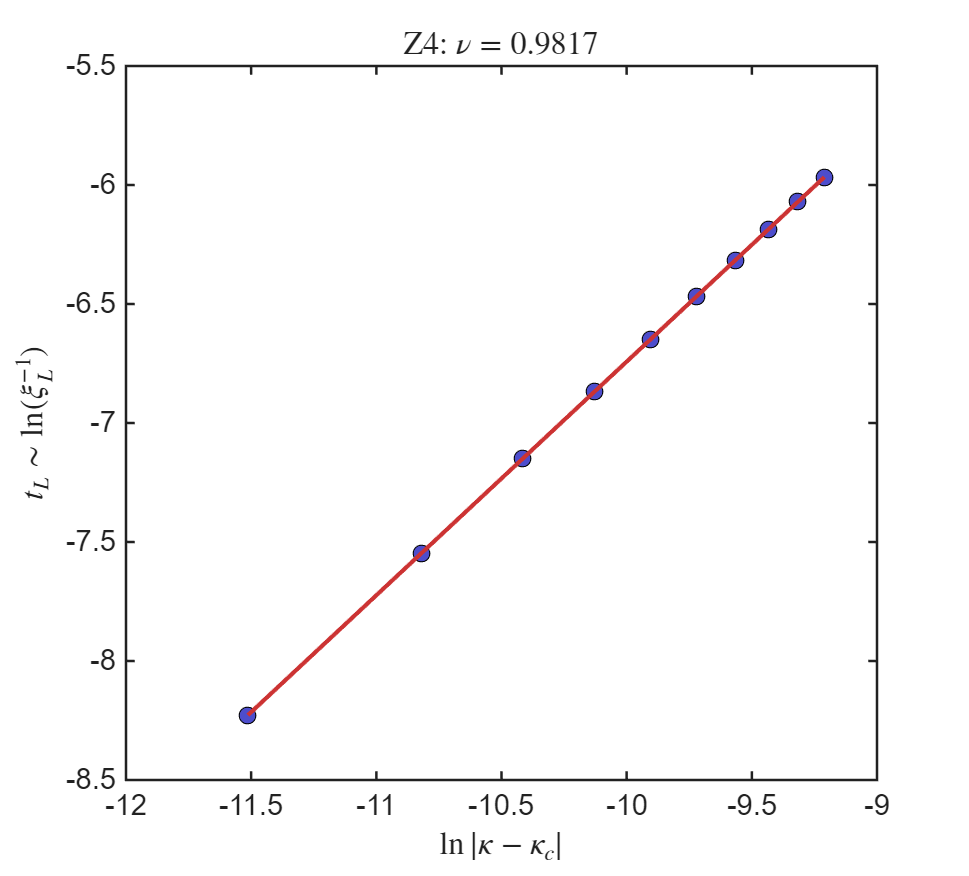} &
        \includegraphics[width=0.31\textwidth]{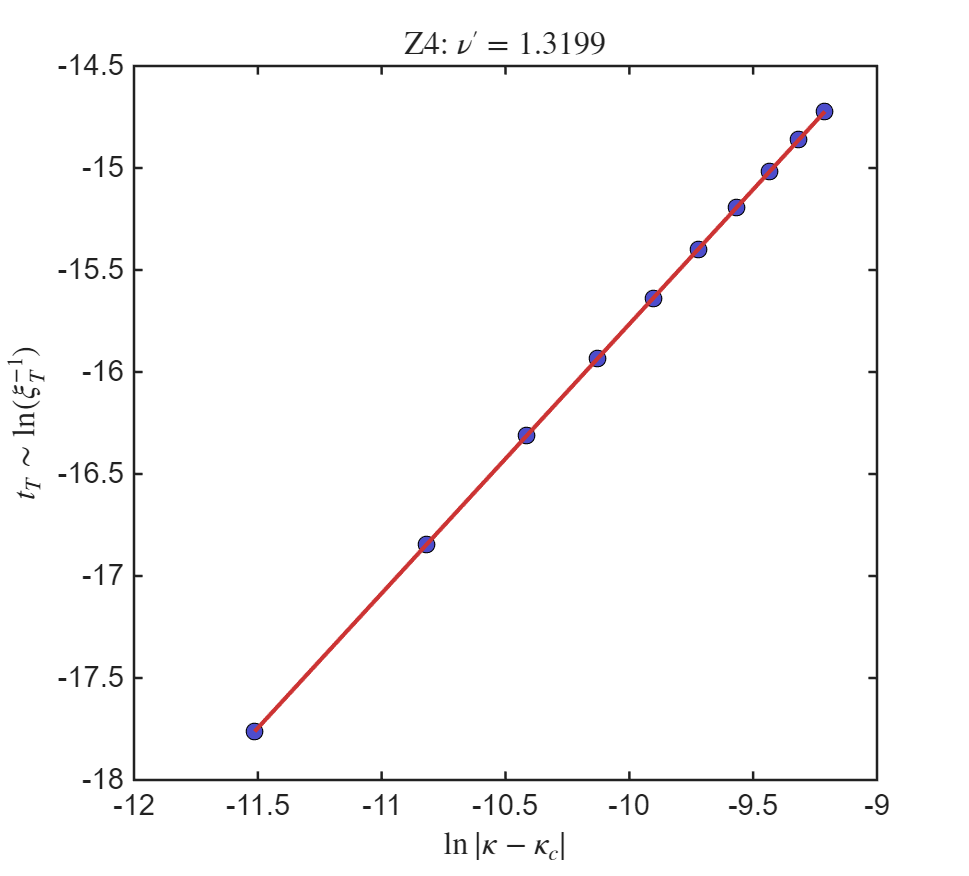} &
        \includegraphics[width=0.31\textwidth]{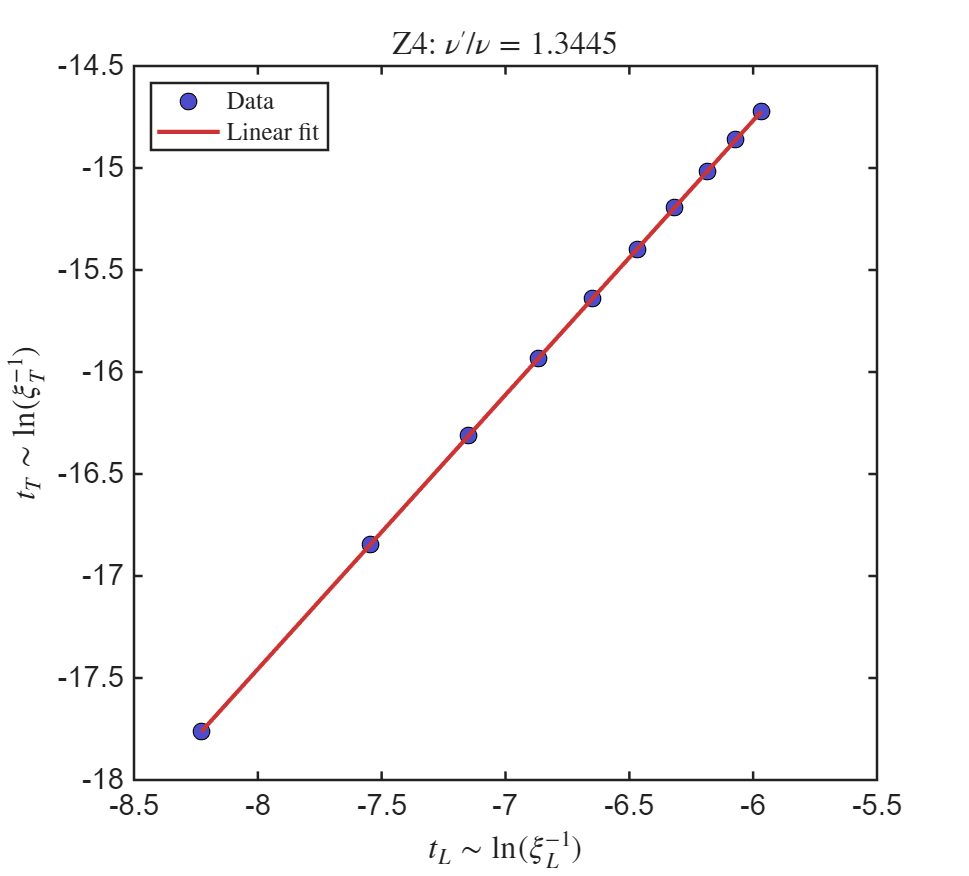} \\
        (a) & (b) & (c)
    \end{tabular}
    \caption{Extraction of $\nu$, $\nu'$ and $\nu'/\nu$ for the $Z_4$ model from linear fits.
(a) $t_L$ against $\ln|\kappa-\kappa_c|$, whose slope gives $\nu$.
(b) $t_T$ against $\ln|\kappa-\kappa_c|$, whose slope gives $\nu'$.
(c) $t_T$ against $t_L$, whose slope gives $\nu'/\nu$.
The first near-critical trajectory shown in Fig.~\ref{fig:rg-flows}(d) is excluded from the fits, and the remaining ten off-critical trajectories are used.}
    \label{fig:nihe}
\end{figure*}

As illustrated in Fig.~\ref{fig:rg-flows}(d), we fix $\Lambda_{01}$ to a small nonzero value and tune the remaining couplings to the SUSY fixed point. Varying $\Delta\kappa=\kappa-\kappa_c$ on the ordered side generates a family of RG trajectories from which the two RG times $t_L$ and $t_T$ are extracted. Since these RG times differ from the logarithms of the inverse length scales only by nonuniversal additive constants, their slopes with respect to $\ln|\Delta\kappa|$ determine the corresponding critical exponents. The fitting procedure is shown in Fig.~\ref{fig:nihe}.

For the $Z_4$ analysis, we use initial $\kappa=\kappa_c+\delta\kappa$ with $\kappa_c\simeq0.009196935$, $\delta\kappa$ sampled between $0$ and $10^{-4}$, and fixed initial anisotropy $\Lambda_{01}=10^{-5}$. The linear behavior in Fig.~\ref{fig:nihe} supports the extraction of $\nu$, $\nu'$, and $\nu'/\nu$ within this fitting window. Since scaling is asymptotic, the window must remain sufficiently close to the critical point; farther away from criticality, nonlinear corrections from irrelevant operators can spoil the linear scaling behavior.

\begin{table}[!htbp]
    \caption{Critical exponents and scaling law comparison at $N_f = 0.5$ with LPA$'$6 truncation in the SSB expansion. The exponents $\nu$, $\nu'$, and $\nu'/\nu$ are extracted from linear fits. The anisotropy eigenvalue $|y_n|$ is obtained from the SSB stability matrix, and the last two rows are calculated directly from $|y_n|$.}
    \label{tab:nup-over-nu}
    \begin{ruledtabular}
    \begin{tabular}{lccc}
        & $Z_4$ & $Z_5$ & $Z_6$ \\
        $\nu$ & $0.9817$ & $0.9817$ & $0.9817$ \\
        $\nu'$ & $1.3199$ & $2.9229$ & $4.1429$ \\
        $\nu'/\nu$ & $1.3445$ & $2.9773$ & $4.2200$ \\
        $|y_n|$ & $0.6889$ & $3.9555$ & $6.4418$ \\
        $1+|y_n|/2$ & $1.3444$ & $2.9778$ & $4.2209$ \\
        $1+|y_n|/3$ & $1.2296$ & $2.3185$ & $3.1473$ \\
    \end{tabular}
    \end{ruledtabular}
\end{table}

The fitted exponents obtained by applying the above procedure are summarized in Table~\ref{tab:nup-over-nu}, together with $|y_n|$ and the corresponding scaling law estimates. Here $|y_n|$ denotes the absolute value of the leading irrelevant eigenvalue associated with the dimensionless anisotropic coupling, obtained from the stability matrix in the SSB expansion---distinct from the SYM-derived value discussed earlier. As a consistency check, the same eigenvalue analysis gives $\nu=0.9818$, in excellent agreement with the values of $\nu$ directly extracted from the linear fits in Table~\ref{tab:nup-over-nu}. At the same time, the RG-flow analysis further determines the second length-scale exponent $\nu'$, which cannot be directly obtained from the stability matrix alone.

The data in Table~\ref{tab:nup-over-nu} allow us to test which scaling law governs the second length scale. The fitted values of $\nu'/\nu$ agree with
\begin{equation}
    \frac{\nu'}{\nu}=1+\frac{|y_n|}{2},
\end{equation}
for the SUSY fixed points of the $Z_4$, $Z_5$, and $Z_6$ models, whereas the alternative form $1+|y_n|/3$ is inconsistent with the fitted ratios.

This comparison clarifies the different behavior of the two length-scale exponents in Table~\ref{tab:nup-over-nu}. The ordinary correlation-length exponent remains nearly unchanged for the three models, reflecting the common critical behavior controlled by the SUSY fixed point. By contrast, the second length-scale exponent is strongly affected by the discrete anisotropy, and its rapid increase follows from the increasing magnitude of the dangerously irrelevant eigenvalue $|y_n|$. The resulting separation between $\nu'$ and $\nu$ is much larger than in the $Z_3$ FIQCP case studied in Ref.~\cite{torresFermioninducedQuantumCriticality2018}, where the small irrelevant eigenvalue leads to only a weak difference between the two length-scale exponents. The large values of $\nu'$ obtained here may therefore provide a more favorable setting for observing the second length scale in future numerical simulations.

\section{Discussion and Conclusion}\label{sec:discussion}

\subsection{\texorpdfstring{Possible Explanation for $p=2$}{Possible Explanation for p=2}}

The ordered-phase analysis above supports the scaling law $\nu'/\nu=1+|y_n|/2$. This result should be compared with Ref.~\cite{patilUnconventionalU1Zq2021}, where the same scaling form was studied in clock models using quantum Monte Carlo (QMC) simulations. In that work, the isotropic three-dimensional classical clock model gives $p=2$, whereas quantum clock models and strongly anisotropic classical clock models give $p=3$.

In this sense, the denominator $p=2$ found here is not unexpected. Although the present model contains gapless fermions and is quantum in origin, the FRG calculation is formulated in Euclidean spacetime with an equal velocity ansatz and an isotropic cutoff. The temporal and spatial directions are therefore treated on the same footing in the present truncation, closer in spirit to an isotropic classical flow than to the strongly anisotropic setting where $p=3$ was observed.

A more decisive test of the $p=3$ scenario would require a formulation that keeps temporal and spatial renormalizations independent. Simply introducing a velocity factor $v$ into the derivative terms is not sufficient if the regulator still has the isotropic form $p_0^2+v^2p_i^2<k^2$, because in that setup $v$ does not acquire a nontrivial flow. A possible route is to develop regulator schemes that treat temporal and spatial momenta separately, although the construction and validation of such a formulation are beyond the scope of the present work.

\subsection{\texorpdfstring{Comparison of Different Estimates of $\nu'$}{Comparison of Different Estimates of nu prime}}

The main estimate of $\nu'$ in this work is obtained directly from the ordered-phase SSB flow. The RG times $t_L$ and $t_T$ are determined from the crossings of the dimensionless longitudinal and transverse masses, and the exponents $\nu$, $\nu'$, and $\nu'/\nu$ are extracted independently from the corresponding linear fits. In this sense, the SSB calculation does not rely on the scaling law as a prior assumption; instead, it provides an independent numerical test of it. This is different from the estimate in the $Z_3$ Dirac problem of Ref.~\cite{torresFermioninducedQuantumCriticality2018}, where $\nu'$ was obtained by combining fixed-point data with the scaling law.

A second estimate can be obtained by combining the SYM fixed-point data with the scaling law verified above. The SYM expansion gives an independent estimate of the critical exponent $\nu$ and the anisotropy eigenvalue $y_n$. Substituting these values into $\nu'/\nu=1+|y_n|/2$ gives Table~\ref{tab:sym-nu-prime-estimate}.
\begin{table}[!htbp]
    \caption{Estimate of $\nu'$ obtained by combining the SYM fixed-point exponents with the scaling law $\nu'/\nu=1+|y_n|/2$.}
    \label{tab:sym-nu-prime-estimate}
    \begin{ruledtabular}
    \begin{tabular}{cccc}
        & $Z_4$ & $Z_5$ & $Z_6$ \\
        $\nu$ & 0.9292 & 0.9292 & 0.9292 \\
        $|y_n|$ & 0.6826 & 4.3428 & 7.0017 \\
        $\nu'$ & 1.2464 & 2.9469 & 4.1822 \\
    \end{tabular}
    \end{ruledtabular}
\end{table}

The SYM-based estimates are close to the direct SSB results in Table~\ref{tab:nup-over-nu}. The two calculations are based on different expansion schemes and projection points, so small quantitative differences are natural. Among the values of $\nu$ and $\nu'$ compared here, the largest relative deviation from the direct SSB estimates is about $5.6\%$.

The distinction between the two expansions suggests how the calculation could be extended to larger $N_f$. In the SYM expansion, a moderately negative mass parameter can indicate the tendency toward symmetry breaking and does not by itself invalidate the fixed-point analysis. In the SSB expansion, however, the expansion point is the running minimum $\kappa$. Once the relevant ordered-phase trajectory cannot be kept in a regime with a positive and well-controlled $\kappa$, the SSB expansion around $(\rho,\tau)=(\kappa,0)$ itself ceases to provide a valid description of the ordered phase. Nevertheless, the phenomenon of two length scales is expected to persist at larger $N_f$, as illustrated for example by the $Z_3$ FIQCP case. A pseudospectral FRG approach, which treats the effective potential more globally in field space, may help address this problem and thus provides a possible direction for future work \cite{borchardtGlobalSolutionsFunctional2015,borchardtSolvingFunctionalFlow2016}.

\subsection{Conclusion}

In this work, we used the functional renormalization group within the LPA$'$ truncation to study $(2+1)$-dimensional Dirac systems with discrete $Z_n$ anisotropy. We focused on two related questions: whether $\mathcal{N}=2$ SUSY can emerge at critical points for $n>3$, and how dangerously irrelevant anisotropy affects the ordered phase through the generation of multiple length scales.

Our fixed-point analysis based on the SYM expansion shows that, for $Z_4$, $Z_5$, and $Z_6$, the leading anisotropic perturbations are irrelevant at the supersymmetric critical point. As a result, the critical point has only one relevant direction and can control a continuous phase transition. The fixed point is consistent with the $(2+1)$-dimensional $\mathcal{N}=2$ Wess-Zumino universality class, as supported by the approximate SUSY coupling relation and by the near equality of the bosonic and fermionic anomalous dimensions. In contrast, emergent SUSY is absent in the $Z_3$ case due to the relevant cubic anisotropy, in agreement with the previous FIQCP studies \cite{liFermioninducedQuantumCritical2017,jianFermioninducedQuantumCritical2017a,torresFermioninducedQuantumCriticality2018}.

In the ordered phase, the SSB expansion makes it possible to track how the dangerously irrelevant anisotropy becomes physically effective in the infrared. Although the anisotropy is irrelevant at the SUSY critical point, it generates a finite transverse mass in the ordered phase and freezes the would-be angular mode. The thresholds $m_L^2=1$ and $m_T^2=1$ then provide operational definitions of two RG scales associated with the two characteristic lengths $\xi$ and $\xi'$. This allows us to extract $\nu$, $\nu'$, and $\nu'/\nu$ directly from the ordered-phase RG flow, rather than assuming the scaling law as an input.

For the $Z_4$, $Z_5$, and $Z_6$ models, the directly fitted exponent ratios agree with the scaling law $\nu'/\nu=1+|y_n|/2$, where $y_n$ is the RG eigenvalue associated with the dangerous anisotropic perturbation. The alternative form $1+|y_n|/3$ is not supported within the present calculation. The appearance of the factor $2$ is consistent with the isotropic Euclidean FRG setup used here, in which temporal and spatial directions are treated on the same footing. Testing whether a $p=3$ law can emerge might require a formulation with independent temporal and spatial renormalization.

Several extensions are natural. First, an anisotropic FRG scheme with separate temporal and spatial regulators would be useful for testing whether the $p=3$ scaling law can arise in a genuinely anisotropic quantum setting. Second, going beyond LPA$'$ by including momentum dependence, nonlocal structures, or additional nonminimal couplings could further clarify the stability of the SUSY fixed point. Third, pseudospectral \cite{borchardtGlobalSolutionsFunctional2015,borchardtSolvingFunctionalFlow2016} or other global-potential FRG methods may help access regimes where the SSB expansion around a well-controlled positive $\kappa$ becomes difficult. Finally, the large values of $\nu'$ found here may be accessible in future quantum Monte Carlo studies, which would provide an independent test of the two-length-scale critical behavior predicted in this work.

\begin{acknowledgments}
This project is supported by the National Natural Science Foundation of China (Grant No. 12222515), the Research Center for Magnetoelectric Physics of Guangdong Province (Grant No. 2024B0303390001), the Guangdong Provincial Key Laboratory of Magnetoelectric Physics and Devices (Grant No. 2022B1212010008), the Science and Technology Projects in Guangzhou City (Grant No. 2025A04J5408), and Quantum Science and Technology-National Science and Technology Major Project (Grant No. 2025ZD0300400). T. Y. Wang is also supported by (national) college students innovation and entrepreneurship training program, Sun Yat-sen University.
\end{acknowledgments}

\appendix
\section{SSB and SYM Expansion}
\subsection{Notation for the SSB Phase}
To perform the expansion, the notation needs to be clarified. 

The bosonic fields are decomposed as
\begin{equation}
    \phi_i = \phi_{i,0} + \Delta\phi_i, \qquad i=1,2,
\end{equation}
where $\phi_{i,0}$ denotes the background value and
$\Delta\phi_i$ denotes the fluctuation.

At the end of the calculation, we will set all fluctuations and fermion fields to zero. We will also choose coordinates such that $\phi_{2,0}=0$. 

We define:
\begin{equation}
    \kappa \equiv \rho_0 = \frac{\phi_{1,0}^2}{2}.
\end{equation}
The minimum values of the fields are given by $\phi_1^{\min} = -\sqrt{2\kappa}$ and $\phi_2^{\min} = 0$.

We adopt the following conventions:
\begin{equation}
    \begin{aligned}
&\Gamma_{k,0}^{(2)} = \left. \Gamma_k^{(2)} \right|_{\Delta\phi = \psi = 0} ,\\
&\Delta\Gamma_k^{(2)} = \Gamma_k^{(2)} - \Gamma_{k,0}^{(2)} ,\\
&\tilde{\partial}_t \quad\text{only acts on the regulator}.
\end{aligned}
\end{equation}

We then expand following the method proposed in  \cite{classenFluctuationinducedContinuousTransition2017}:

\begin{equation}
\begin{aligned}
    \partial_t \Gamma_k = &
\frac{1}{2} \tilde{\partial}_t \mathrm{STr} \ln\left( \Gamma_k^{(2)} + R_k \right) 
\\=& \frac{1}{2} \tilde{\partial}_t \mathrm{STr} \ln\left( \Gamma_{k,0}^{(2)} + R_k \right) 
+ \\
&\frac{1}{2} \tilde{\partial}_t \mathrm{STr} \sum_{n=1}^{\infty} \frac{(-1)^{n+1}}{n} 
\left[ \left( \Gamma_{k,0}^{(2)} + R_k \right)^{-1} \Delta\Gamma_k^{(2)} \right]^n.
\end{aligned}
\label{eq:wetterich_expansion}
\end{equation}

At the minimum of the effective potential, i.e., $\Delta\phi=0$, we have $u_{11}=m_L^2, u_{22}=m_T^2$.

\subsection{\texorpdfstring{Flow Equations for $\lambda$s and $\Lambda$s}{Flow Equations for lambda and Lambda Couplings}}

In the SSB expansion, the flow equations for the vertices of the effective potential can be extracted in this manner:
\begin{equation}\label{eq:Lambdaprotection}
\begin{aligned}
\partial_t \Lambda_{i,j} =
\Biggl[
\partial_t
\frac{\partial^{i+j} u}
{\partial \rho^i \partial \tau^j}
+
\left(
\frac{\partial}{\partial\kappa}
\frac{\partial^{i+j} u}
{\partial \rho^i \partial \tau^j}
\right)\partial_t \kappa
\Biggr]_{\rho=\kappa,\tau=0}.
\end{aligned}
\end{equation}

Substituting Eq.~\eqref{eq:znuflom} into Eq.~\eqref{eq:Lambdaprotection} and denoting $\Sigma_B = \frac{1}{1+m_L^2}+\frac{1}{1+m_T^2}$, $\Sigma_F = \frac{1}{1+2h^2\rho}$, we obtain:
\begin{equation}\label{partialLambda}
    \begin{aligned}
        \partial_t \Lambda_{i,j} =& \left[-D + \frac{1}{2}(D-2+\eta_\phi)(2i+nj)\right]\Lambda_{i,j}\\
        &+\frac{4v_D}{D} \left(1 - \frac{\eta_\phi}{D+2}\right)\frac{\partial^{i+j}}{\partial \rho^i \partial \tau^j}\Sigma_B \\
        &-  \frac{4v_D N_f d_\gamma}{D} \left(1 - \frac{\eta_\psi}{D+1}\right)\frac{\partial^{i+j}}{\partial \rho^i \partial \tau^j}\Sigma_F\\
         &+ \Lambda_{i+1,j}\partial_t \kappa.
    \end{aligned}
\end{equation}

Using $\Lambda_{1,0}=0$, we obtain $\partial_t \kappa = -\frac{\partial_t u^{(1,0)}}{\Lambda_{2,0}}$, so that
\begin{equation}\label{partialkappa}
    \begin{aligned}
        \partial_t \kappa=& (2-D-\eta_\phi)\kappa - \frac{1}{\Lambda_{2,0}}\left[\frac{4v_D}{D} \left(1 - \frac{\eta_\phi}{D+2}\right)\frac{\partial \Sigma_B}{\partial \rho}\right.\\
 &\left.-  \frac{4v_D N_f d_\gamma}{D} \left(1 - \frac{\eta_\psi}{D+1}\right)\frac{\partial \Sigma_F}{\partial \rho}\right].
    \end{aligned}
\end{equation}

In the SYM phase expansion, the flow equations are:
\begin{equation}
    \partial_t \lambda_{i,j} = \left. \left[\partial_t\frac{\partial^{i+j}}{\partial \rho^i \partial \tau^j}  u(\rho,\tau)\right] \right|_{\rho=0,\tau=0}.
\end{equation}

Substituting Eq.~\eqref{eq:znuflom}, calculating with fluctuations included, and finally setting $m_L^2=m_T^2=m^2,\rho=0$, yields:
\begin{equation}\label{partiallambda}
    \begin{aligned}
        \partial_t \lambda_{i,j} =& \left[-D + \frac{1}{2}(D-2+\eta_\phi)(2i+nj)\right]\lambda_{i,j}\\
        &+\left[\frac{4v_D}{D} \left(1 - \frac{\eta_\phi}{D+2}\right)\frac{\partial^{i+j}}{\partial \rho^i \partial \tau^j}\Sigma_B \right.\\
        &\left.-  \frac{4v_D N_f d_\gamma}{D} \left(1 - \frac{\eta_\psi}{D+1}\right)\frac{\partial^{i+j}}{\partial \rho^i \partial \tau^j}\Sigma_F\right]_{L=T,\rho = 0}.
    \end{aligned}
\end{equation}

\section{Threshold Functions}

Threshold functions are integral components of the FRG flow equations, arising from the momentum integration of the regulator's scale derivative $\partial_t R_k$ with the regulated propagators. They encode the non-perturbative content of the quantum fluctuations. In this work, we employ the Litim regulator \cite{LITIM2002128, litimOptimisedRenormalisationGroup2001, litimMindGap2001, litimOptimisationExactRenormalisation2000}. The specific definitions and their evaluated forms, which are based on the results of Ref.~\cite{torresFermioninducedQuantumCriticality2018}, are given below.

\begin{widetext}
We define $\omega_\psi = 2h^2\kappa$.

\begin{equation}
l_{0}^{(B_{1/2})}=\frac{1}{4v_{D}}\int\frac{d^{D}p}{(2\pi)^{D}}p^{2}\frac{\partial_{t}r_{\phi}-\eta_{\phi}r_{\phi}}{p^{2}(1+r_{\phi})+k^{2}m_{L/T}^2},
\end{equation}

\begin{equation}
l_{0}^{(F)}=\frac{1}{2v_{D}}\int\frac{d^{D}p}{(2\pi)^{D}}p^{2}\frac{(1+r_{\psi})(\partial_{t}r_{\psi}-\eta_{\psi}r_{\psi})}{p^{2}(1+r_{\psi})^{2}+k^{2}\omega_\psi},
\end{equation}

\begin{equation}
l_{nm}^{(FB_{1/2})}=-\frac{1}{4v_{D}}k^{2(n+m)-D}\tilde{\partial}_{t}\int\frac{d^{D}p}{(2\pi)^{D}}\frac{1}{(p^{2}(1+r_{\psi})^{2}+k^{2}\omega_{\psi})^{n}(p^{2}(1+r_{\phi})+k^{2}m_{L/T}^2)^{m}},
\end{equation}

\begin{equation}
\begin{aligned}
    l_{111}^{(FB_1B_2)}=-\frac{1}{4v_{D}}k^{6-D}\tilde{\partial}_{t}\int\frac{d^{D}p}{(2\pi)^{D}}\frac{1}{(p^{2}(1+r_{\psi})^{2}+k^{2}\omega_{\psi})(p^{2}(1+r_{\phi})+k^{2}m_L^2)(p^{2}(1+r_{\phi})+k^{2}m_T^2)},
\end{aligned}
\end{equation}

\begin{equation}
m_{4}^{(B_{1/2})}=-\frac{1}{4v_{D}}k^{6-D}\tilde{\partial}_{t}\int\frac{d^{D}p}{(2\pi)^{D}}p^{2}\left(\frac{\partial}{\partial p^{2}}\frac{1}{p^{2}(1+r_{\phi})+k^{2}m_{L/T}^2}\right)^{2},
\end{equation}

\begin{equation}
m_{22}^{(B_1B_2)}=-\frac{1}{4v_{D}}k^{6-D}\tilde{\partial}_{t}\int\frac{d^{D}p}{(2\pi)^{D}}p^{2}\left(\frac{\partial}{\partial p^{2}}\frac{1}{p^{2}(1+r_{\phi})+k^{2}m_L^2}\right)\left(\frac{\partial}{\partial p^{2}}\frac{1}{p^{2}(1+r_{\phi})+k^{2}m_T^2}\right),
\end{equation}

\begin{equation}
m_{4}^{(F)}=-\frac{1}{4v_{D}}k^{4-D}\tilde{\partial}_{t}\int\frac{d^{D}p}{(2\pi)^{D}}p^{4}\left(\frac{\partial}{\partial p^{2}}\frac{1+r_{\psi}}{p^{2}(1+r_{\psi})^{2}+k^{2}\omega_\psi}\right)^{2},
\end{equation}

\begin{equation}
m_{2}^{(F)}=-\frac{1}{4v_{D}}k^{6-D}\tilde{\partial}_{t}\int\frac{d^{D}p}{(2\pi)^{D}}p^{2}\left(\frac{\partial}{\partial p^{2}}\frac{1}{p^{2}(1+r_{\psi})^{2}+k^{2}\omega_\psi}\right)^{2},
\end{equation}

\begin{equation}
m_{12}^{(FB_{1/2})}=-\frac{1}{4v_{D}}k^{4-D}\tilde{\partial}_{t}\int\frac{d^{D}p}{(2\pi)^{D}}p^{2}\frac{1+r_{\psi}}{p^{2}(1+r_{\psi})^{2}+k^{2}\omega_{\psi}}\frac{\partial}{\partial p^{2}}\frac{1}{p^{2}(1+r_{\phi})+k^{2}m_{L/T}^2}.
\end{equation}

Performing the integrals yields

\begin{equation}
l_{0}^{(B_{1/2})}=\frac{2}{D}\left(1-\frac{\eta_{\phi}}{D+2}\right)\frac{1}{1+m_{L/T}^2},
\end{equation}

\begin{equation}
l_{0}^{(F)}=\frac{2}{D}\left(1-\frac{\eta_{\psi}}{D+1}\right)\frac{1}{1+\omega_\psi},
\end{equation}

\begin{equation}
l_{nm}^{(FB_{1/2})}=\frac{2}{D}\left[\left(1-\frac{\eta_{\psi}}{D+1}\right)\frac{1}{1+\omega_{\psi}}+\left(1-\frac{\eta_{\phi}}{D+2}\right)\frac{1}{1+m_{L/T}^2}\right]\frac{1}{(1+\omega_{\psi})^{n}(1+m_{L/T}^2)^{m}},
\end{equation}

\begin{equation}
l_{111}^{(FB_1B_2)}=\frac{2}{D}\left[\left(1-\frac{\eta_{\psi}}{D+1}\right)\frac{1}{1+\omega_{\psi}}+\left(1-\frac{\eta_{\phi}}{D+2}\right)(\frac{1}{1+m_L^2}+\frac{1}{1+m_T^2})\right]\frac{1}{(1+\omega_{\psi})(1+m_L^2)(1+m_T^2)},
\end{equation}

\begin{equation}
m_{4}^{(B_{1/2})}=\frac{1}{(1+m_{L/T}^2)^{4}},
\end{equation}

\begin{equation}
m_{22}^{(B_1B_2)}=\frac{1}{(1+m_L^2)^{2}(1+m_T^2)^{2}},
\end{equation}

\begin{equation}
m_{4}^{(F)}=\frac{1}{(1+\omega_\psi)^{4}}+\frac{1-\eta_{\psi}}{D-2}\frac{1}{(1+\omega_\psi)^{3}}-\left(\frac{1-\eta_{\psi}}{2D-4}+\frac{1}{4}\right)\frac{1}{(1+\omega_\psi)^{2}},
\end{equation}

\begin{equation}
m_{2}^{(F)}=\frac{1}{(1+\omega_\psi)^{4}},
\end{equation}

\begin{equation}
m_{12}^{(FB_{1/2})}=\left(1-\frac{\eta_{\phi}}{D+1}\right)\frac{1}{(1+\omega_{\psi})(1+m_{L/T}^2)^{2}}.
\end{equation}

\end{widetext}

\onecolumngrid
\section{Additional Ordered-Phase RG Flows and Fits}
\label{app:ordered-extra}

This appendix collects the ordered-phase RG flow and fitting figures for the $Z_5$ and $Z_6$ models, supplementing the representative $Z_4$ results shown in the main text.

\begin{figure}[!hbp]
    \centering
    \begin{tabular}{@{}cc@{}}
        \includegraphics[width=0.47\textwidth]{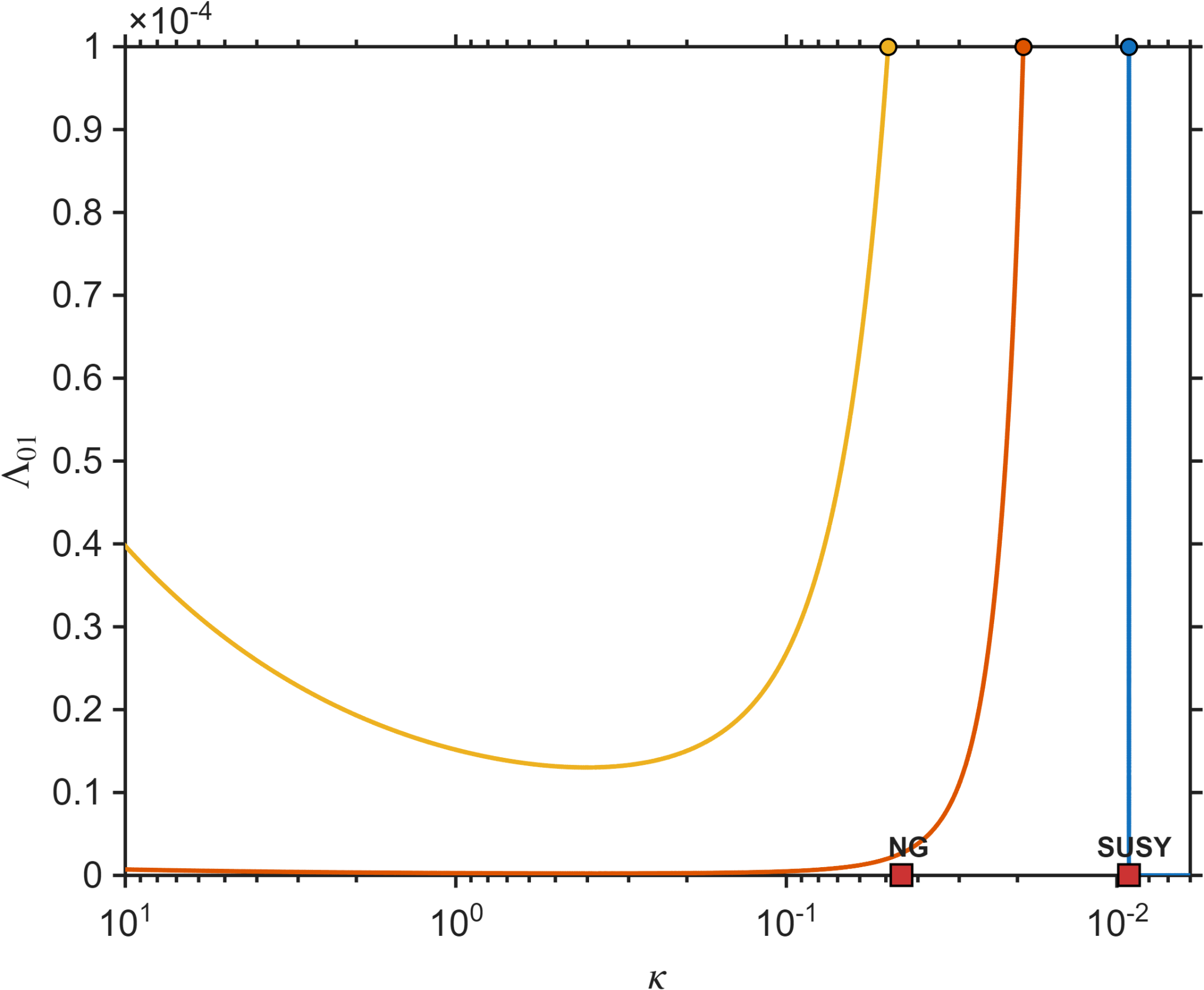} &
        \includegraphics[width=0.47\textwidth]{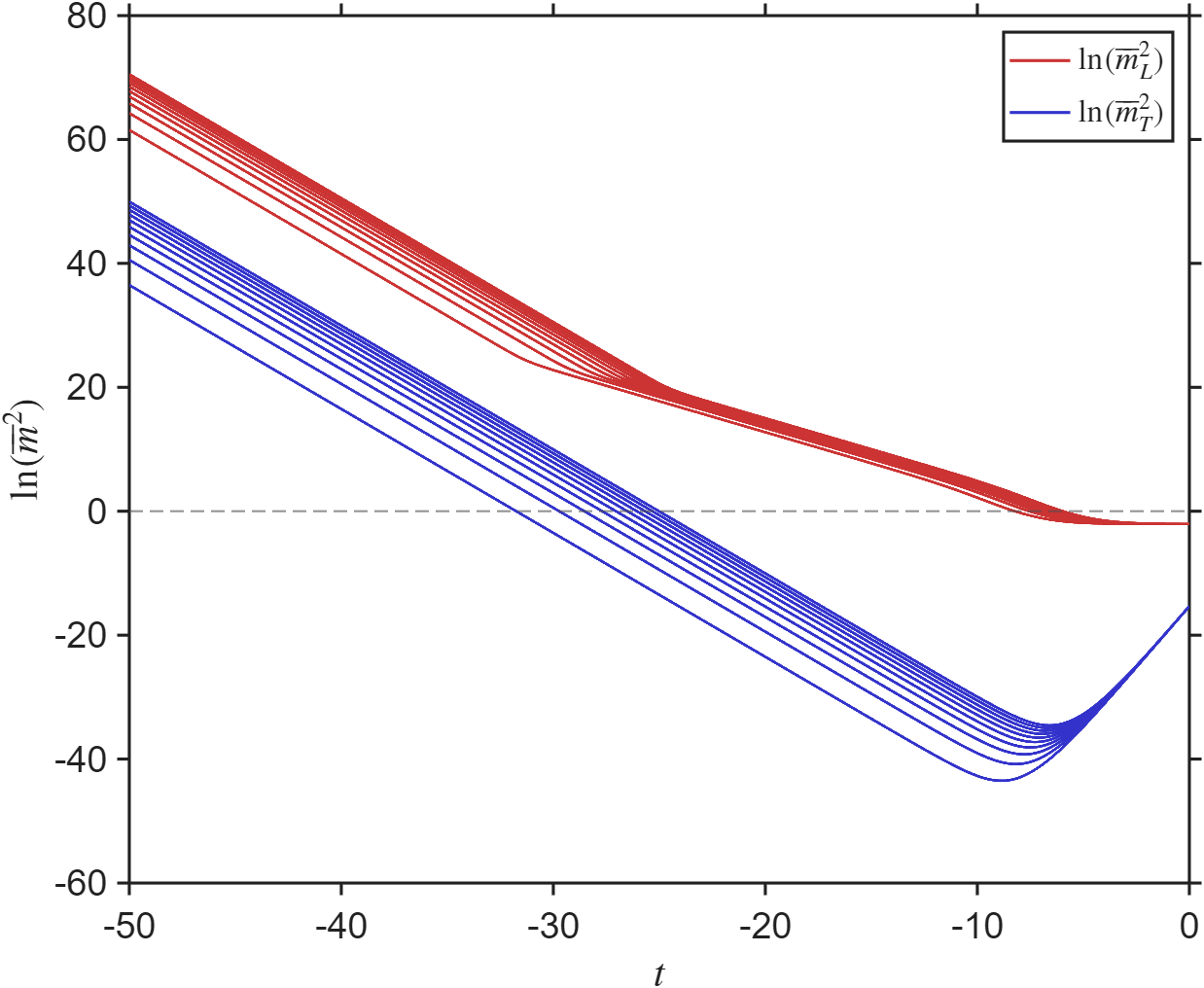} \\
        (a) & (b)
    \end{tabular}
    \caption{Ordered-phase RG flows for the $Z_5$ model at $N_f=0.5$.
(a) Phase portrait in the $(\kappa,\Lambda_{01})$ plane.
(b) Multi-trajectory dimensionless mass flows for different initial values of $\kappa$, used to determine $t_L$ and $t_T$ from the crossings $\ln m_L^2=0$ and $\ln m_T^2=0$.}
    \label{fig:app-ordered-rg-z5}
\end{figure}

\begin{figure}[!htbp]
    \centering
    \begin{tabular}{@{}cc@{}}
        \includegraphics[width=0.47\textwidth]{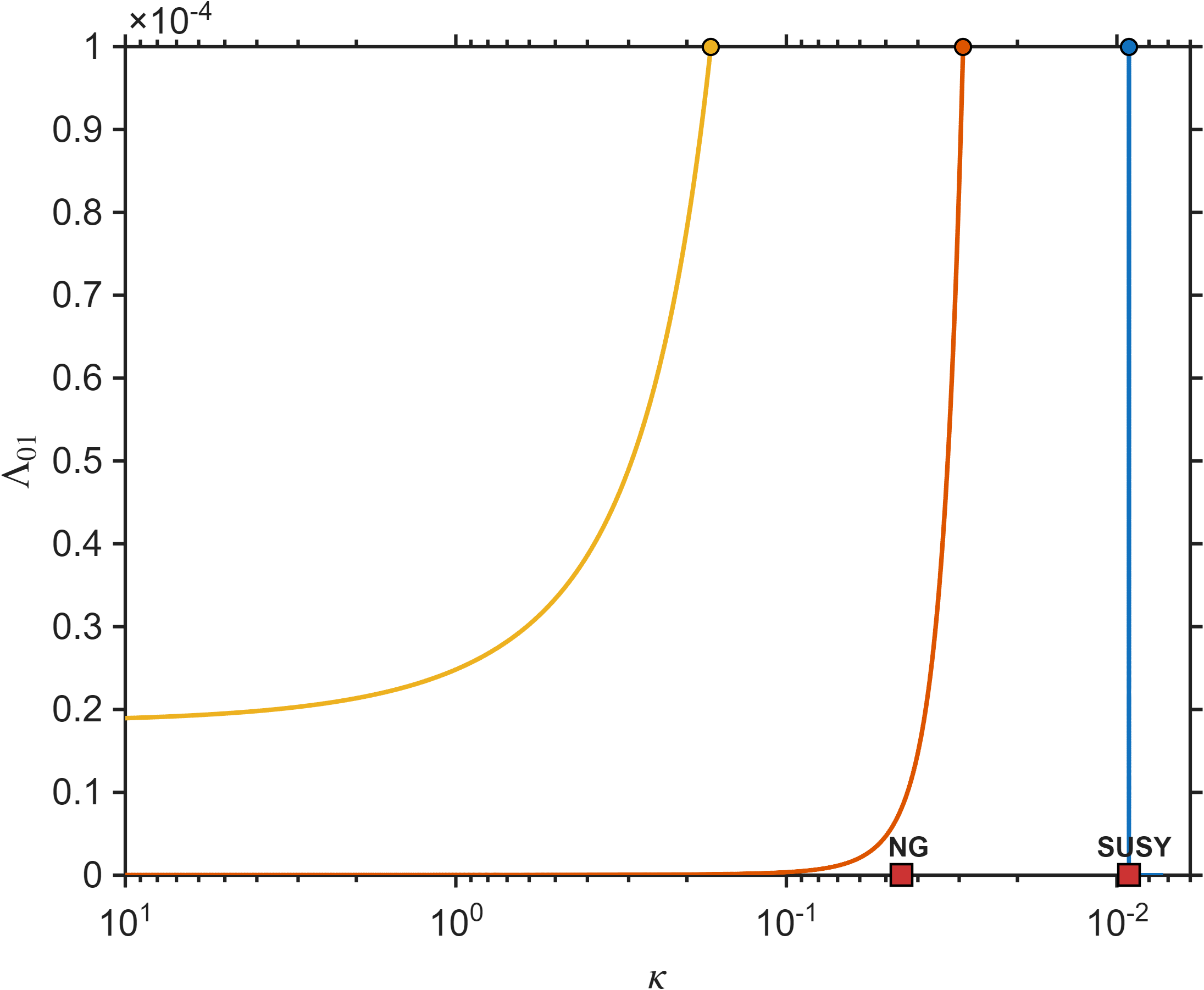} &
        \includegraphics[width=0.47\textwidth]{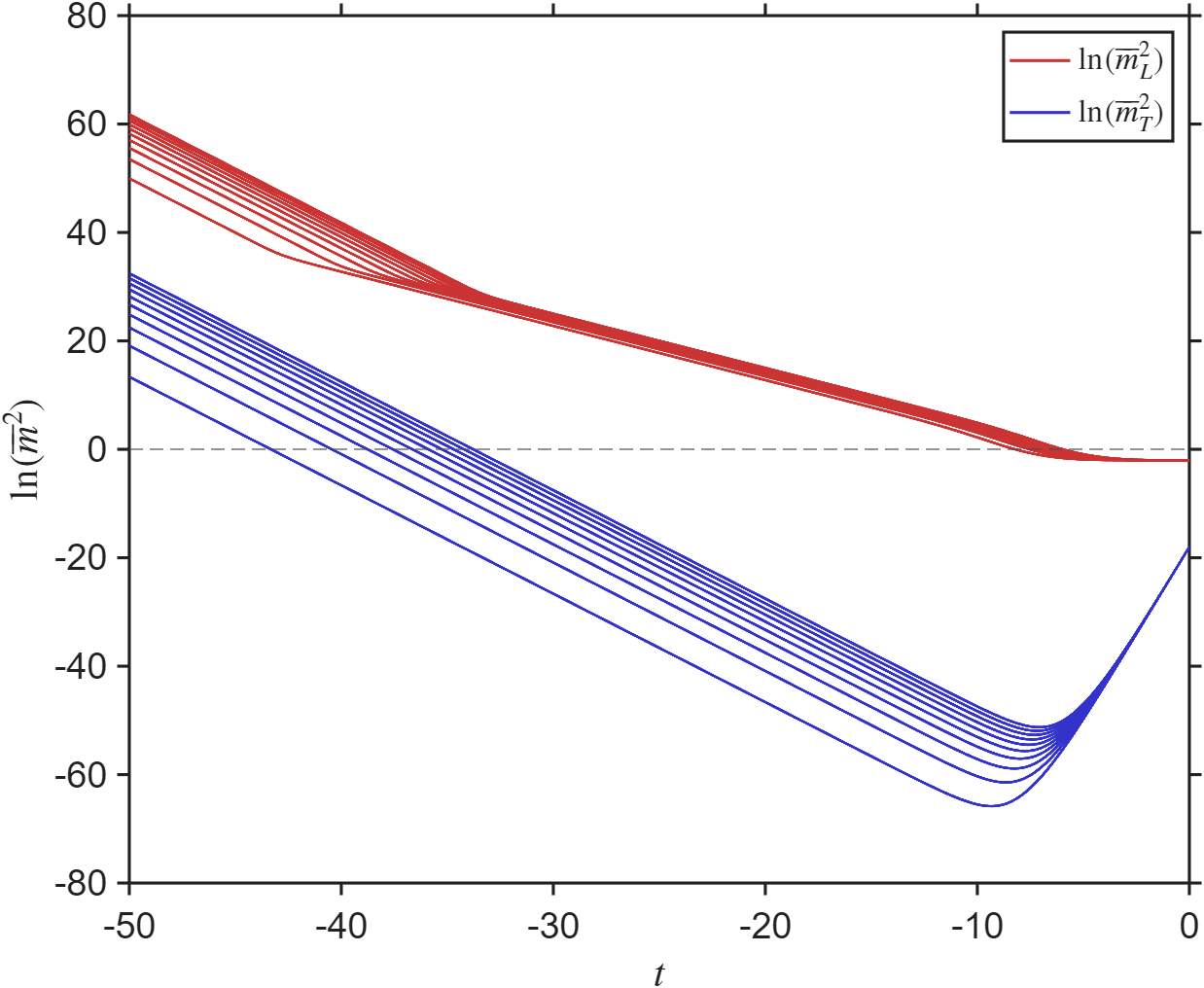} \\
        (a) & (b)
    \end{tabular}
    \caption{Ordered-phase RG flows for the $Z_6$ model at $N_f=0.5$.
(a) Phase portrait in the $(\kappa,\Lambda_{01})$ plane.
(b) Multi-trajectory dimensionless mass flows for different initial values of $\kappa$, used to determine $t_L$ and $t_T$ from the crossings $\ln m_L^2=0$ and $\ln m_T^2=0$.}
    \label{fig:app-ordered-rg-z6}
\end{figure}

\begin{figure}[!htbp]
    \centering
    \begin{tabular}{@{}ccc@{}}
        \includegraphics[width=0.315\textwidth]{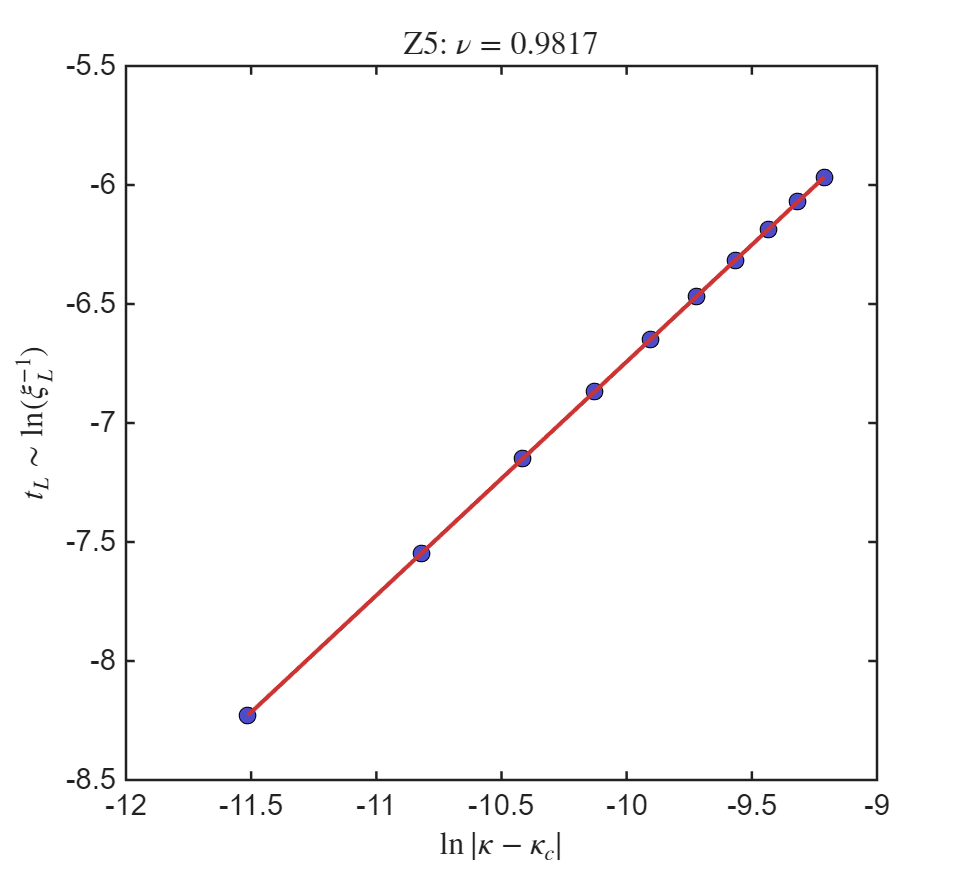} &
        \includegraphics[width=0.315\textwidth]{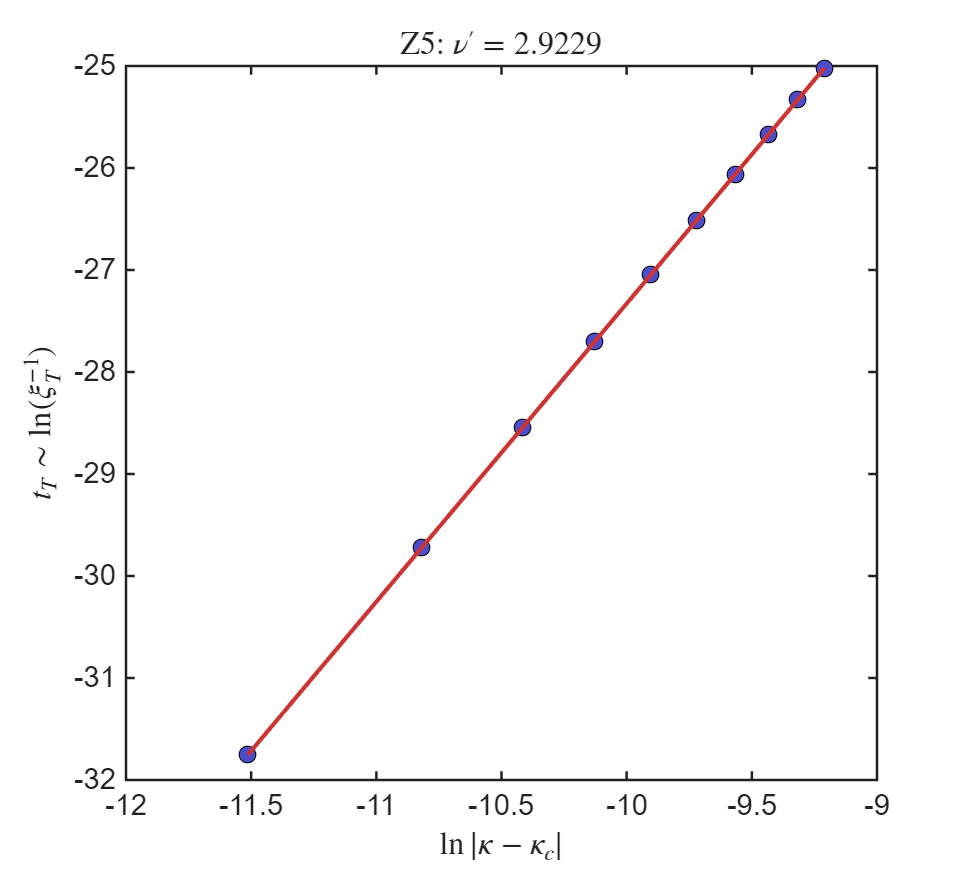} &
        \includegraphics[width=0.315\textwidth]{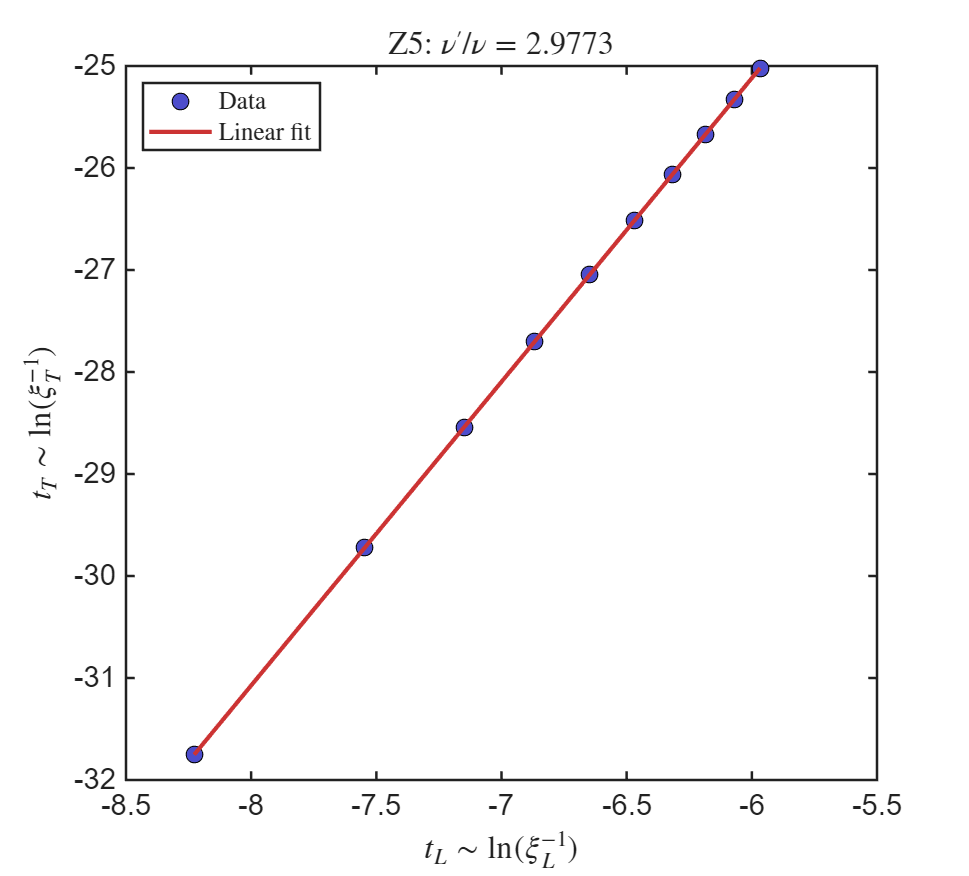} \\
        (a) & (b) & (c)
    \end{tabular}
    \caption{Extraction of $\nu$, $\nu'$ and $\nu'/\nu$ for the $Z_5$ model from linear fits.
(a) $t_L$ against $\ln|\kappa-\kappa_c|$.
(b) $t_T$ against $\ln|\kappa-\kappa_c|$.
(c) $t_T$ against $t_L$.}
    \label{fig:app-ordered-fits-z5}
\end{figure}

\begin{figure}[!htbp]
    \centering
    \begin{tabular}{@{}ccc@{}}
        \includegraphics[width=0.315\textwidth]{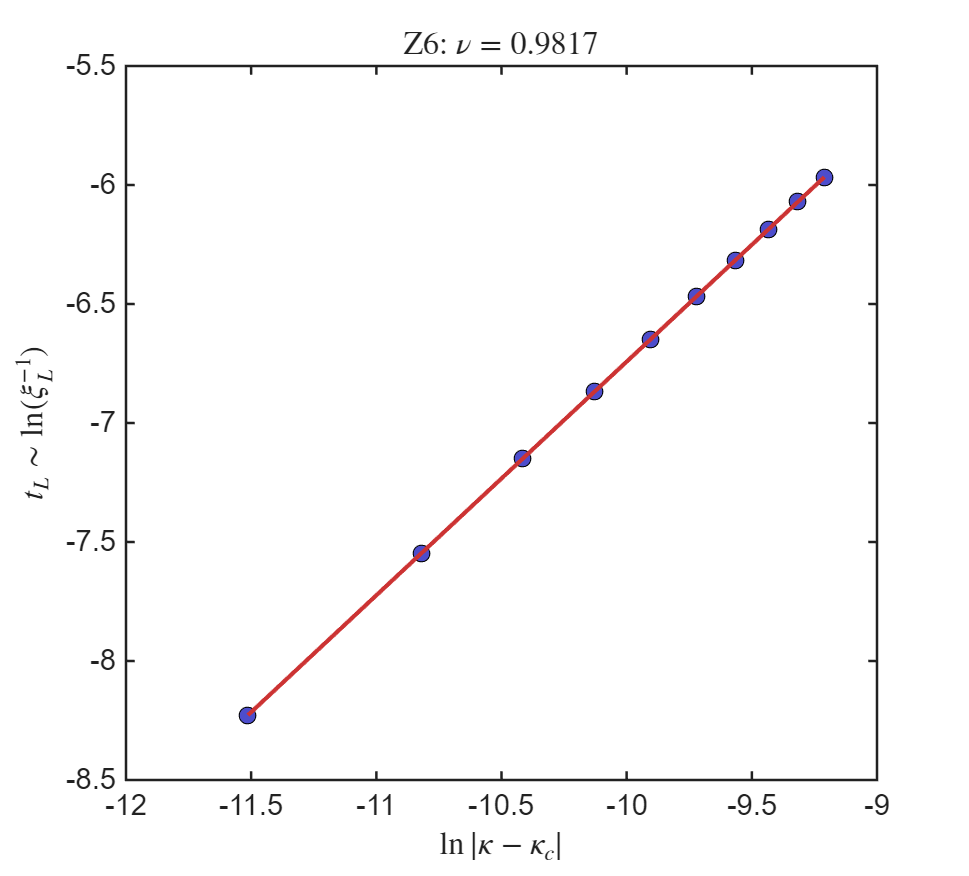} &
        \includegraphics[width=0.315\textwidth]{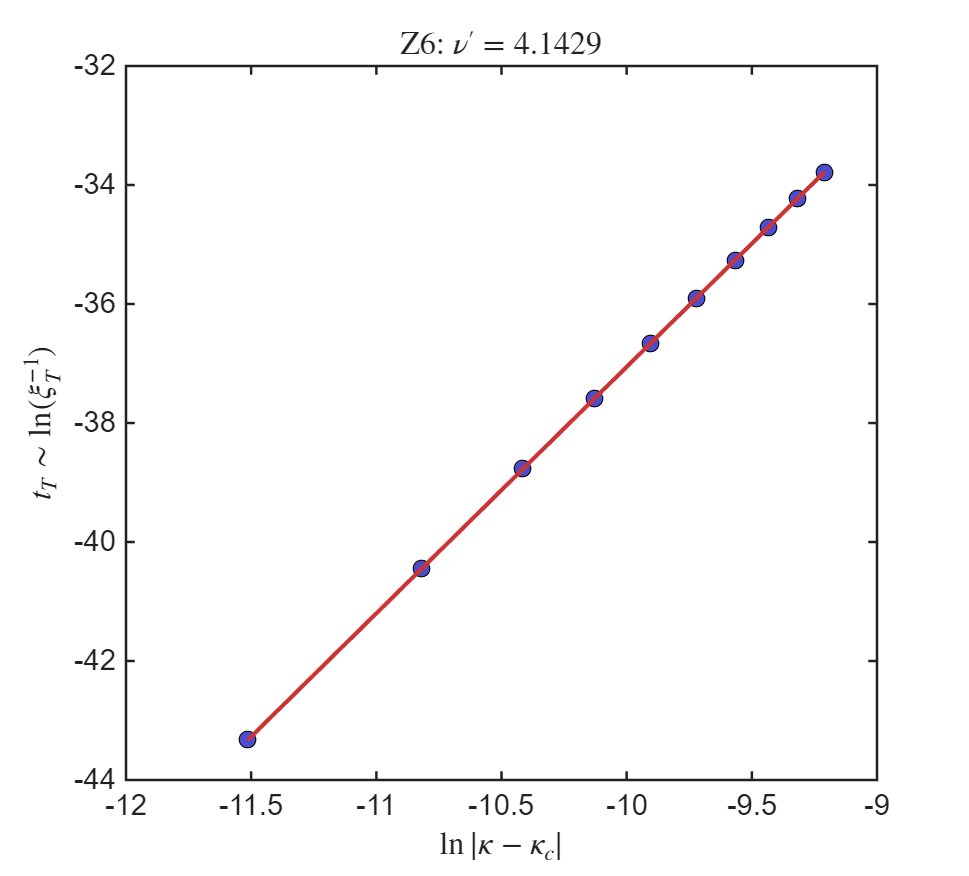} &
        \includegraphics[width=0.315\textwidth]{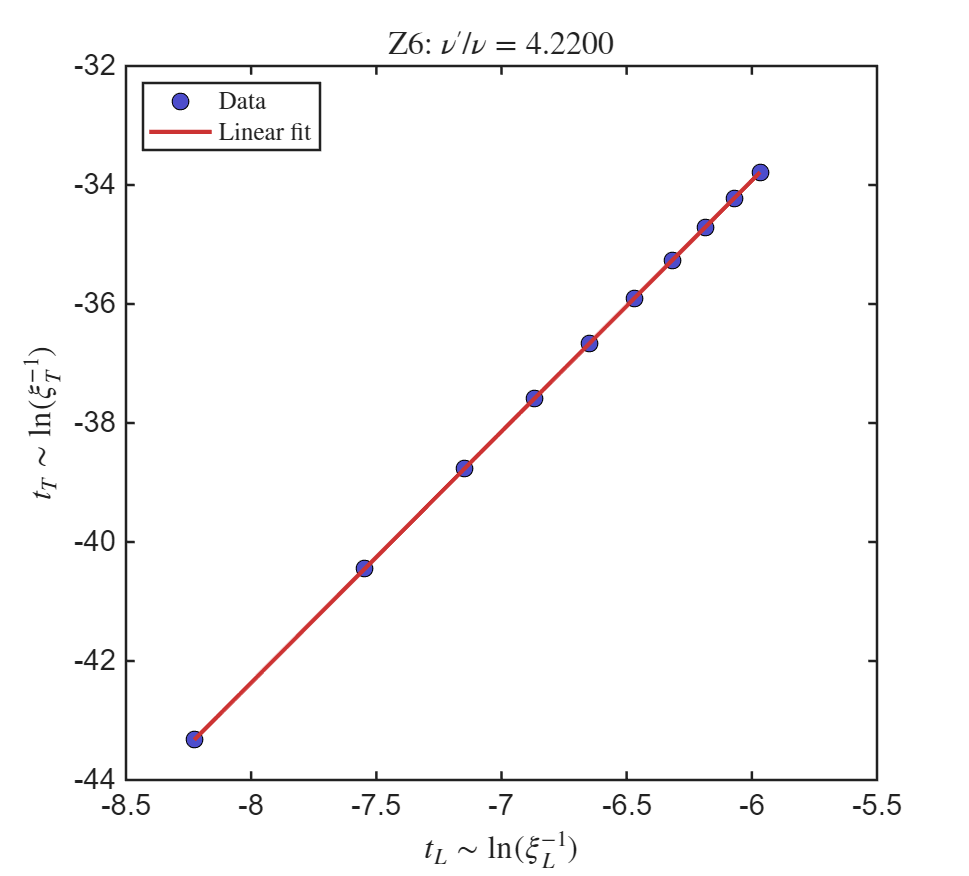} \\
        (a) & (b) & (c)
    \end{tabular}
    \caption{Extraction of $\nu$, $\nu'$ and $\nu'/\nu$ for the $Z_6$ model from linear fits.
(a) $t_L$ against $\ln|\kappa-\kappa_c|$.
(b) $t_T$ against $\ln|\kappa-\kappa_c|$.
(c) $t_T$ against $t_L$.}
    \label{fig:app-ordered-fits-z6}
\end{figure}

The same protocol as in the main text is used for both models. We fix the initial anisotropy to $\Lambda_{01}=10^{-5}$ and choose initial values $\kappa=\kappa_c+\delta\kappa$, where $\delta\kappa$ takes 11 equally spaced values between $0$ and $10^{-4}$. The critical values are $\kappa_c=0.009196795351803$ for $Z_5$ and $\kappa_c=0.009196789000928$ for $Z_6$. In each case, the first near-critical trajectory with $\delta\kappa=0$ is shown in the RG-flow plots, Figs.~\ref{fig:app-ordered-rg-z5}(b) and \ref{fig:app-ordered-rg-z6}(b), but is excluded from the fits; the remaining 10 off-critical trajectories are used to extract $\nu$, $\nu'$, and $\nu'/\nu$. The corresponding fitting plots are shown in Figs.~\ref{fig:app-ordered-fits-z5} and \ref{fig:app-ordered-fits-z6}.

Figs.~\ref{fig:app-ordered-rg-z5}(a) and \ref{fig:app-ordered-rg-z6}(a) show the ordered-side flow in the $(\kappa,\Lambda_{01})$ plane. The $Z_5$ flow still displays a visible growth of $\Lambda_{01}$ toward the infrared, whereas in the $Z_6$ case $\Lambda_{01}$ approaches an approximately constant value. This behavior follows from the canonical part of the ordered-phase flow. In the deep ordered regime, where fluctuation corrections are suppressed and $\eta_\phi\to0$, the anisotropic coupling obeys
\begin{equation}
    \partial_t\Lambda_{01}
    =
    \left[
        \frac{n}{2}(D-2+\eta_\phi)-D
    \right]\Lambda_{01}
    \xrightarrow[D=3,\ \eta_\phi\to0]{}
    \left(\frac{n}{2}-3\right)\Lambda_{01}.
\end{equation}
Since $t=\ln(k/\Lambda)$ decreases toward the infrared, this canonical scaling implies an infrared growth of $\Lambda_{01}$ for $Z_4$ and $Z_5$, while the $Z_6$ anisotropy is marginal at the canonical level. The nearly constant $\Lambda_{01}$ in the $Z_6$ flow therefore does not imply the absence of anisotropic effects. As shown in Eq.~\eqref{eq:ordered-mass-scaling}, the transverse mass is controlled by the combined factor $n^2\kappa^{n/2-1}\Lambda_{01}$, so it can continue to grow even when $\Lambda_{01}$ itself is nearly constant.

\twocolumngrid

\bibliography{SUSYcriticality}

\end{document}